\documentclass[pra,aps,twocolumn,showpacs]{revtex4}

\usepackage[english]{babel}
\usepackage{multirow}
\usepackage{pifont}                           % AMS math symbols.
\usepackage{amssymb}                           % AMS math symbols.
\usepackage{amsmath}                           % AMS extensions.
\usepackage{graphicx}                          % Include figure files.
\usepackage{bm}                                % For bold math.
\usepackage{color}
\usepackage{psfrag}
\usepackage{fix-cm}
\newcommand{\etal}[0]{{\it et al.}}              % Quantum operator.

\begin{document}

%%
%%---- Title of the paper --------------------------------------------------------------------------------
%%
\title{Quantitative study of quasi-one-dimensional Bose gas experiments via the stochastic Gross-Pitaevskii equation}

%%
%%---- Authors and affiliations --------------------------------------------------------------------------
%%
\author{S.\ P.\ Cockburn}%\email[E-mail: ]{s.p.cockburn@ncl.ac.uk}
\author{D.\ Gallucci}
\author{N.\ P.\ Proukakis}

\affiliation{School of Mathematics and Statistics, Newcastle University, Newcastle upon Tyne, 
NE1 7RU, United Kingdom\\
}

%\date{Fri 21st Jan 2011}

%%
%%---- Abstract ----------------------------------------------------------------------
%%

\begin{abstract}
The stochastic Gross-Pitaevskii equation is shown
to be an excellent model for quasi-one-dimensional Bose gas experiments,
accurately reproducing the {\it in situ} density profiles recently obtained in
the experiments of Trebbia {\it et al.} [Phys. Rev. Lett. {\bf 97}, 250403 (2006)]
and van Amerongen {\it et al.} [Phys. Rev. Lett. {\bf 100}, 090402 (2008)], 
and the density fluctuation data
reported by Armijo {\it et al.} [Phys. Rev. Lett. {\bf 105}, 230402 (2010)].
To facilitate such agreement, we propose and implement a quasi-one-dimensional stochastic 
equation for the low-energy, axial modes, while atoms in excited transverse modes 
are treated as independent ideal Bose gases.

\end{abstract}

%%
%%---- PACS numbers 
%%
% \pacs{03.75.Nt; 03.75.Lm; 05.10.Gg}
%67.85.-d}
%03.75.Hh - Static properties of condensates; thermodynamical, statistical, and structural properties;
%67.85.-d - Ultracold gases, trapped gases (see also 03.75.-b Matter waves in quantum mechanics);
%67.85.Bc;
%05.10.Gg;
%42.50.-p
\maketitle

\section{Introduction}

%General:
Ultracold atomic gases are 
proving to be extremely useful 
tools for synthesizing low dimensional quantum models
owing to the huge degree of controlability they offer \cite{Bloch2008}.
The effective system dimensionality may be tuned in experiments 
by manipulation of external trapping potentials,
with the underlying physics ultimately
set by the trap geometry and level of 
quantum degeneracy \cite{Gorlitz2001}. %,Koehl2005(more refs? atom chip reviews?. 
Increasing the trapping potential in one direction leads 
to an effectively two-dimensional (2D) system 
\cite{Gorlitz2001,Rychtarik2004,Stock2005,Smith2005}, 
which allows access to a number of interesting phenomena such as the 
Berezinskii-Kosterlitz-Thouless transition and related studies on
the nature of the Bose gas in two-dimensions
\cite{Hadzibabic2006,Schweikhard2007,Kruger2007,Rath2008,Clade2009,Rath2010,Tung2010,Hung2011}.
Increasing the trapping potential in
a further direction %two directions 
results instead in an effectively one-dimensional (1D) system 
\cite{Moritz2003,Paredes2004,Kinoshita2004,Cacciapuoti2003,Trebbia2006,vanAmerongen2008,Armijo2010,Armijo2011,Esteve2006,
Dettmer2001,Hellweg2003,Gerbier2003,Richard2003,Manz2010,
Gustavson1997,Hinds2001,Schumm2005,Hofferberth2006,Fixler2007,Jo2007b,Hofferberth2007,Gross2010,Baumgartner2010,
Hinds1999,Ott2001,Hansel2001,Folman2002,Fortagh2007,Atomchips2011}.
%\cite{Gustavson1997,Hinds1999,Hinds2001,Dettmer2001,Hansel2001,Ott2001,Folman2002,
%Hellweg2003,Gerbier2003,Richard2003,
%Cacciapuoti2003,Paredes2004,Kinoshita2004,Schumm2005,Esteve2006,Trebbia2006,
%Fortagh2007,Fixler2007,
%Jo2007b,vanAmerongen2008,Manz2010,Armijo2010,Gross2010,Baumgartner2010,Atomchips2011}.

%Gustavson1997,Hinds2001,Schumm2005,Fixler2007,Jo2007b,Hofferberth2007,Gross2010,Baumgartner2010
%Hinds1999,Hansel2001,Ott2001,Folman2002,Fortagh2007,Atomchips2011

%1D regimes:
In a 1D set-up, 
one may obtain \cite{Petrov2000} either a weakly-interacting system, or, 
for rather low densities, 
a strongly-interacting Tonks-Girardeau gas \cite{Girardeau1960,Moritz2003,Paredes2004,Kinoshita2004}.
The finite temperature phase diagram of a {\em weakly interacting} 1D Bose 
gas \cite{Petrov2000,AlKhawaja2003,Armijo2011} is more complex than that of a 3D
gas, due to a separation in the temperatures for the onset 
of density and phase fluctuations. 
Density fluctuations are typically
suppressed at higher temperatures than phase fluctuations, allowing
for the formation of a so-called quasi-condensate \cite{PopovBook}. 
\begin{figure}[b!]
  \centerline{
		\psfrag{yaxis}{\scalebox{3.75}{$k_{B}T/\hbar\omega_{\perp}$}}
    \psfrag{xaxis}{\scalebox{3.75}{$\mu/\hbar\omega_{\perp}$}}
    \includegraphics[angle=0,scale=0.3,clip]{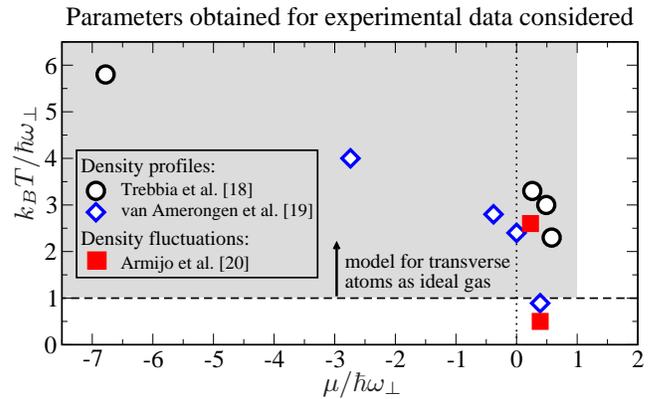}
		%figures/mu_T_plane_v2_newExp_HFparams.eps}%mu3d_T_plane_v5.eps}
  }
  \caption[]{
    (Color online) Phase diagram indicating the theoretically obtained parameters
    for the experiments considered. Hollow symbols indicate density profile
    data of \cite{Trebbia2006} and \cite{vanAmerongen2008}, 
    whereas filled symbols indicate the density fluctuation data of 
    \cite{Armijo2010}.
  }
  \label{fig:one}
\end{figure}
A number of experiments have recently probed the physics
%regimes of quantum degeneracy 
of highly-elongated finite temperature Bose gases
at equilibrium,
including the direct observation and analysis of 
density \cite{Cacciapuoti2003,Esteve2006,Armijo2010} and phase fluctuations 
\cite{Dettmer2001,Hellweg2003,Gerbier2003,Richard2003,Manz2010}.
Understanding the
properties of matter waves in such geometries 
is of key importance to
%a fundamental precursor to their use in
% important as they are
%ideal candidates 
%an important component in
%potential new devices %for precision measurements
%such as 
atom interferometers
\cite{Gustavson1997,Hinds2001,Schumm2005,Fixler2007,Jo2007b,Hofferberth2007,Gross2010,Baumgartner2010} 
and atom chips \cite{Hinds1999,Hansel2001,Ott2001,Folman2002,Hofferberth2006,Fortagh2007,Atomchips2011}.

%Motivate our model:
%Given the central role played by fluctuations in 
%a finite temperature quasi-1D Bose gas, it is crucial to include these
%effects in order to give an accurate theoretical description of the entire 
%weakly interacting regime. Moreover, while the essential physics 
%within such tightly confined geometries may be 
%governed by 1D effective theories, real experiments are carried out 
%within three-dimensions. 
%For certain parameters, therefore, the deviation 
%away from a true-1D system becomes non-negligible and
%is apparent in experimental data. 

In this paper, we show that {\em in situ} density profiles and density fluctuations
from the elongated Bose gas experiments 
of Trebbia \etal \cite{Trebbia2006},
van Amerongen \etal \cite{vanAmerongen2008}, and 
Armijo \etal \cite{Armijo2010}, can be predicted {\em ab initio}
by means of an effective 1D stochastic model;
to achieve this, we propose and implement for the first time a modification
to the usual form of the stochastic Gross-Pitaevskii equation 
\cite{Stoof1999,Stoof2001,Gardiner2003}
which additionally incorporates quasi-1D effects.
Such an extension beyond the purely 1D limit is required, since 
these experiments probe regimes
for which both $\mu,\, k_{B} T \gtrsim \hbar \omega_{\perp}$,
as evident from Fig.\ref{fig:one}. Thus, one should in general
%
%modified to incorporate quasi-1D effects. 
%% and 
%%which we refer to as the quasi-1D SGPE,
%This is proposed and implemented for the first time in the present work.
%Fig.\ref{fig:one} shows clearly that these experiments probe regimes
%for which both $\mu,\, k_{B} T \gtrsim \hbar \omega_{\perp}$,
%for which an extension beyond the purely 
%1D limit is required.
%In general,
%one should 
account for both a quasi-1D degenerate system exhibiting fluctuations
and for the non-negligible role of transverse thermal modes.
These are included here
by (i) implementing a stochastic quasi-1D equation of state for the 
axial modes, and (ii) treating atoms in excited transverse modes 
as independent, ideal Bose gases, as proposed in a related study
%the latter of which was also used in 
\cite{vanAmerongen2008}.

In Sec.~II we discuss our methodology in more detail, while Sec.~III focuses on
the {\it ab initio} reproduction of the experimental density profiles and 
density fluctuation results reported in 
\cite{Trebbia2006,vanAmerongen2008,Armijo2010}, 
with our conclusions presented in Sec.~IV.

%consider a quantitative test of 
%a model based upon a one-dimensional stochastic 
%Gross-Pitaevskii equation (SGPE) \cite{Stoof1999,Gardiner2003} 
%by comparing to {\it in situ} data 
%from three experiments:
%density profiles reported by Trebbia \etal \cite{Trebbia2006} 
%and van Amerongen \etal \cite{vanAmerongen2008}, and 
%data on density fluctuations from the recent experiment of
%Armijo \etal \cite{Armijo2010}.
%Fig.\ref{fig:one} shows the parameters which we find provide a good 
%match to the data from the three experiments that we consider for comparison.
%It is clear that the data points extend into 
%regimes for which both $\mu,\, k_{B} T \gtrsim 1$, 
%indicating the importance of an extension beyond the purely 
%1D limit of the SGPE. As will be detailed in the following Section, 
%we account for this by using a quasi-one-dimensional equation of state for the 
%axial modes represented by the SGPE, while atoms in excited transverse modes 
%are treated as independent, ideal Bose gases
%\cite{vanAmerongen2008}.

\section{Methodology}
\label{sec:method}

We seek a model which can provide {\it ab initio} predictions for
experimentally measurable properties
obtained by {\it in situ} absorption imaging. 
There are two issues that need to be addressed simultaneously here,
regarding the spatial extent of the quasi-condensate in the
transverse direction, and the role of atoms in excited transverse modes.
As mentioned, the important parameters affecting these are
the ratio of the chemical potential, $\mu$, and 
thermal energy, $k_{B}T$, to the transverse ground state
energy $\hbar\omega_{\perp}$. For $\mu\ll\hbar\omega_{\perp}$, 
the transverse ground state density has a Gaussian profile
of width $l_{\perp}$, the transverse oscillator length,
whereas for larger $\mu$,
this width becomes increased due to the 
effect of repulsive interactions.
%e ground state density 
%until the density becomes more parabolic as the transverse 
%Thomas-Fermi regime is reached.
%as in the Thomas-Fermi solution to the GPE.
On the other hand, if $k_{B}T\ll\hbar\omega_{\perp}$, 
then thermal occupation of the
transverse excited modes is negligibly small; for higher 
temperatures, this is no longer true and atoms in these 
modes will contribute considerably to experimental observables,
such as density profiles. 

In the present work, we treat the 
%axial and transverse 
quasi-condensate and thermal modes separately:
%using a combination
%of different approaches, as discussed below.
consideration of transverse thermal modes is crucial for
matching total atom numbers and density profiles, 
however they may be simply and accurately described as independent
equilibrium Bose gases, as shown in \cite{vanAmerongen2008};
fluctuating axial modes
%whereas phase fluctuations 
are instead treated 
within our modified stochastic model,
a novel feature of this work,
which is discussed in detail below.

{\em Quasi-condensate Modeling -}
The axial modes of a sufficiently elongated Bose gas are typically
subject to phase and density fluctuations due to thermal excitations with
wavelengths greater than the radial extent of the gas
\cite{Stringari1998,Petrov2001}. This 
requires the inclusion of such fluctuations when describing
modes with energies
less than $\hbar\omega_{\perp}$. We therefore choose to describe these axial modes dynamically
using a 
stochastic Gross-Pitaevskii equation (SGPE) \cite{Stoof1999,Stoof2001,Gardiner2003}. 
In solving this equation, we make the following two assumptions:
(i) thermal modes (with energies 
$>\hbar\omega_{\perp}$) may be treated as though at equilibrium, 
and therefore represent a heat bath in contact with the axial sub-system 
%(with energies $<\hbar\omega_{\perp}$).
(these two components are assumed to be in diffusive and thermal 
equilibrium with a temperature $T$ and chemical potential $\mu$); 
(ii) the modes in the weakly trapped,
axial direction are sufficiently highly occupied that the classical
field approximation is valid 
\cite{Svistunov1991,Stoof2001,Duine2001,Davis2001,Sinatra2001,Goral2001}. 

Under assumptions (i) and (ii), the axial modes may be 
represented by the 1D SGPE,
\begin{equation}
  \begin{split}
    i\hbar\frac{\partial \psi(z,t)}{\partial t}& = (1-i\gamma(z,t))
    \bigg[-\frac{\hbar^{2}}{2m}\frac{\partial^{2}}{\partial z^{2}}+V(z)\\
    &+g|\psi|^{2}-\mu\bigg]\psi(z,t)+\eta(z,t),
  \end{split}
  \label{eq:1dSGPE}
\end{equation}
where $\psi$ is a complex order parameter, $V(z)=m\omega_{z}^{2}z^{2}/2$ 
is the axial trapping potential, $g=2\hbar\omega_{\perp}a$
is the one-dimensional interaction strength (with $a$ the s-wave scattering length),
and $\eta$ is a complex Gaussian noise term, with correlations given by the relation
$\langle \eta^{*}(z,t) \eta(z',t') \rangle =2\hbar\gamma(z,t)k_{B}T\delta(z-z')\delta(t-t')$. 
The strength of the noise, and damping, due to contact with the transverse thermal 
modes, is given by $\gamma(z,t)$. This may be calculated {\it ab initio} in terms of the
Keldysh self-energy \cite{Stoof1999,Stoof2001,Duine2001}, however as we are interested only 
in the system properties at equilibrium, we may approximate this quantity to be spatially and 
temporally constant
\footnote{
Although not relevant for the equilibrium quantities probed here, 
perturbations away from equilibrium would require a more 
accurate calculation of $\gamma(z,t)$ including both spatial 
\cite{Duine2001,Cockburn2010} and temporal dependence.
While a time-dependent thermal cloud has yet to be implemented within the SGPE
formalism, both temporal and spatial dependence of the damping parameter 
are calculated fully self-consistently within a complementary  `ZNG' scheme \cite{Zaremba1999},
whose validity is
however restricted to regimes of relatively small phase fluctuations.}.
To a good approximation, this is given by $\gamma = 3 \times 4ma^2 k_{B} T /(\pi\hbar^2)$ 
\cite{Penckwitt2002,Duine2004}.

Eq.\eqref{eq:1dSGPE} is valid in the scenario that the transverse 
ground state is a Gaussian of width $l_{\perp}$. However, 
as we wish 
to consider experimental data from actual quasi-1D 
systems, we should additionally modify
our model
in a manner which accounts for the transverse swelling of 
the Bose gas due to repulsive interactions, 
as seen experimentally \cite{Kruger2010}.
In the context of the ordinary Gross-Pitaevskii equation, 
one may replace the 1D equation of state
$\mu[n]=g n$, where $n$ denotes the density, 
with %that for a quasi-1D quasi-condensate
\cite{Fuchs2003}
\begin{equation}
  \mu[n]=\hbar\omega_{\perp}\left[\sqrt{1+4na}-1\right],
  \label{eq:q1d_eos}
\end{equation}
which is obtained variationally by minimising the 3D GPE 
with respect to the transverse chemical potential 
\cite{Gerbier2004,Mateo2007,Frantzeskakis2010}.
This was shown in \cite{Gerbier2004} to interpolate smoothly across
the 1D-to-3D crossover \cite{Menotti2002}, 
%with 
%Eq.\eqref{eq:q1dSGPE} reducing to
%Eq.\eqref{eq:1dSGPE} 
and reduces to the 1D result
in the limit that $4an\ll1$,
as pointed out (for the ordinary GPE) 
in \cite{Fuchs2003,Mateo2007,Frantzeskakis2010}.

Motivated by this we propose, somewhat heuristically, 
a similar amendment to the 1D stochastic equation (Eq.\eqref{eq:1dSGPE});
for elongated but not truly 1D atomic clouds, this 
gives rise to the modified stochastic equation
%
%We insert this new equation of state to give a modified form for the SGPE, which may be written as
\begin{equation}
  \begin{split}
    i\hbar&\frac{\partial \psi(z,t)}{\partial t} = (1-i\gamma(z,t))
    \bigg[-\frac{\hbar^{2}}{2m}\frac{\partial^{2} }{\partial z^{2}}+V(z)\\
    &+\hbar\omega_{\perp}\left(\sqrt{1+4a|\psi|^{2}}-1\right)-\mu\bigg]\psi(z,t)+\eta(z,t)\;,
  \end{split}
  \label{eq:q1dSGPE}
\end{equation}
which we henceforth refer to as the quasi-1D SGPE.
The proposed modification to the 1D SGPE
is likely to become important when the inequality 
$\mu\ll\hbar\omega_{\perp}$ is no longer satisfied, 
signaling the onset of quasi-1D effects.
This method is of course only intended for modeling
very elongated systems with $\mu \lesssim{\rm few}~\hbar \omega_{\perp}$, in the
weakly-interacting
regime $mg/\hbar^2 n\ll1$; in this work we 
thus restrict the application of this equation to the
weakly interacting regime, where numerous experiments exist.

In our stochastic scheme, the equilibrium state is reached in a dynamical manner,
when the effects of the noise 
and damping terms of Eq.\eqref{eq:q1dSGPE} balance out. 
Although 
we demonstrate that this leads to 
accurate equilibrium predictions, 
the validity of this equation for 
describing dynamical features remains to be investigated.

{\em Thermal Transverse Modes:}
Atoms in the transverse modes are considered to be in 
static equilibrium, distributed according to 
Bose-Einstein statistics; the relative importance of their contributions
%Atoms in these modes will contribute 
%to experimental findings to a degree which is 
depends on
the ratio $k_B T/\hbar\omega_{\perp}$,
and their contribution is significant in most experimentally relevant cases
\cite{Trebbia2006,vanAmerongen2008,Armijo2010}
(except at extremely low temperatures).
To account for their contribution to
total linear density profiles, we compute 
(making the $\mu$ and $T$ dependence explicit for clarity)
\begin{equation}
  n(z;\mu,T)=\langle|\psi(z;\mu,T)|^2\rangle+n_{\perp}(z;\mu,T)
  \label{eq:ntot}
\end{equation}
where
\begin{equation}
  n_{\perp}(z;\mu,T)=\frac{1}{\lambda_{\rm dB}}\sum_{j=1}^{\infty} (j+1) 
  {\rm g}_{1/2}\left[e^{(\mu-V(z)-j\hbar\omega_{\perp})/k_{B}T}\right],
  \label{eq:nperp}
\end{equation}
${\rm g}_{1/2}[\ldots]$ is the polylogarithm (or Bose function) 
of order $1/2$,
and $\lambda_{\rm dB}=h/\sqrt{2\pi m k_{B} T}$ is the 
thermal de Broglie wavelength. 

%The model that we employ follows the procedure given in \cite{vanAmerongen2008},
%with the SGPE, suitably modified to model the dimensional crossover regime,
%replacing the one-dimensional Yang-Yang theory used in the analysis of
Such an addition of particles in thermal modes has already been used in
the so-called modified Yang-Yang model of
\cite{vanAmerongen2008}, where Eq.\ \eqref{eq:nperp} was used
in conjunction with a description of
atoms within axial modes based on Yang-Yang theory.
In our approach, we effectively replace the Yang-Yang theory with the SGPE, 
thereby also providing an indirect comparison between SGPE and Yang-Yang, 
in the weakly-interacting limit.
Furthermore, on making this replacement, the contribution given by $n_{\perp}$ remains 
consistent with our treatment of bosons in excited transverse modes as having already 
reached thermal equilibrium, as assumed within the SGPE model we apply here.
\begin{figure*}[ht]
  \centering
  \begin{tabular}{c}
  \includegraphics[angle=0,scale=0.375,clip]{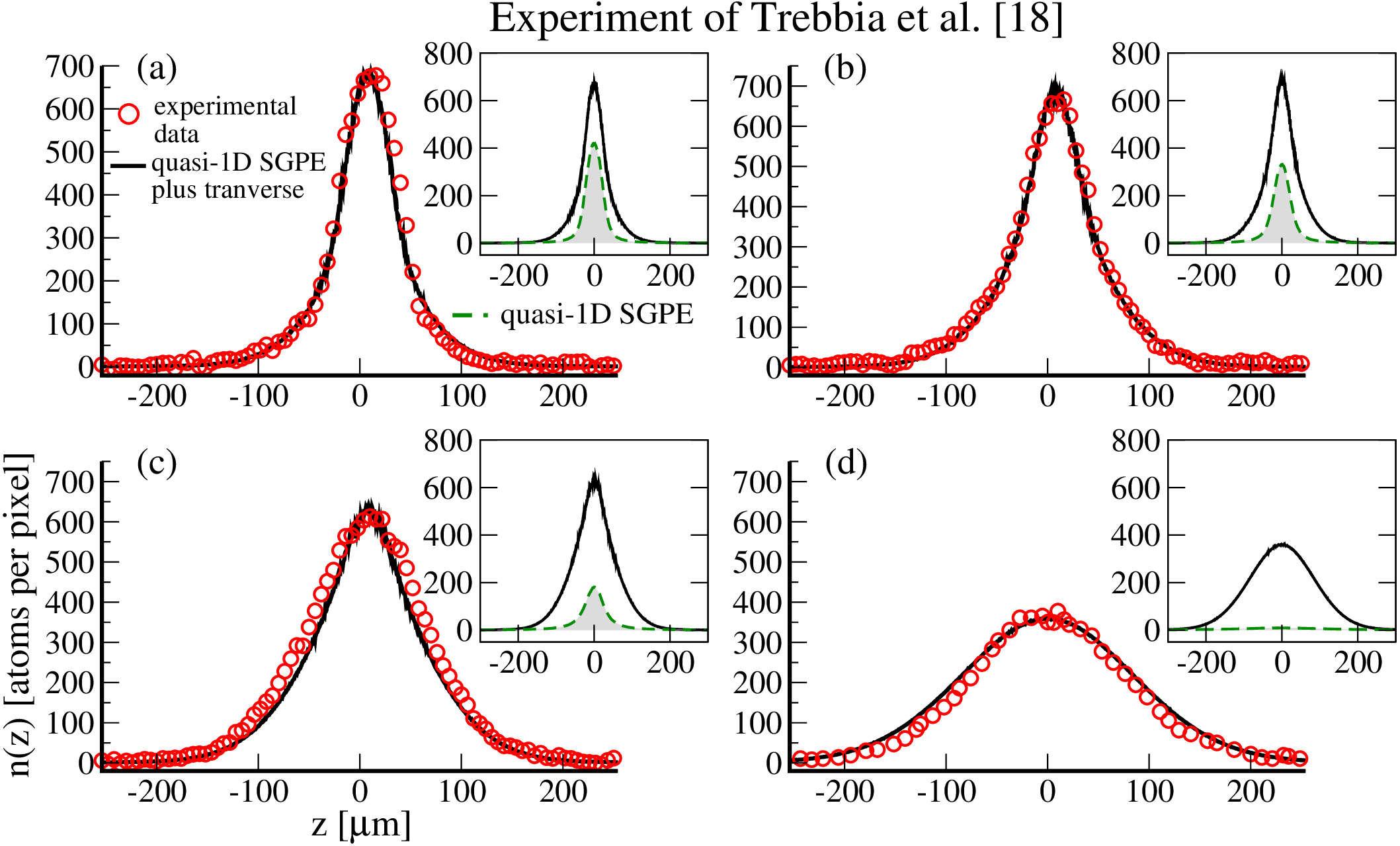}\\					
      \begin{tabular}{ c | c || c | c |}
      %Row 1
				 &  &
  	  	\multicolumn{2}{|c|}{$\mu/\hbar\omega_{\perp}$} \\
			\hline
      %Row 2
    		& \multirow{2}{*}{$k_{B}T/\hbar\omega_{\perp}$} & Analysis of \cite{Trebbia2006}  
   		  & \multicolumn{1}{|c|}{Current analysis} \\
        \cline{3-4}
      %Row 3
        &  & 3D Hartree-Fock & Quasi-1D SGPE \\
      \hline 
      \hline
      %Row 4 
      \multirow{1}{*}{Fig. $2$a} 
		  	& 2.3 & 0.4 & 0.58 \\
	      \cline{2-4}
      \hline
      %Row 5
      \multirow{1}{*}{Fig. $2$b} 
		  	& 3.0 & 0.65 & 0.49 \\
	      \cline{2-4}
      \hline
      %Row 6
      \multirow{1}{*}{Fig. $2$c} 
		  	& 3.3 & 0.6 & 0.26 \\
	      \cline{2-4}
      \hline
      %Row 7
      \multirow{1}{*}{Fig. $2$d} 
		  	& 5.8 & -6.78 & -6.78 \\
	      \cline{2-4}
      \hline
    \end{tabular}
  \label{table:HFF_mus}

  \end{tabular}
  \caption[]{
    (Color online) Total linear density profiles from the quasi-1D SGPE model 
		(Eq.\eqref{eq:ntot}; black solid line) 
    versus data from the experiment of Trebbia \etal \cite{Trebbia2006} (red circles).
		%\footnote{The data in Figs.~\ref{fig:HFF}(a-c) are from the experiment, while 
		%the data in Fig.~\ref{fig:HFF}(d)	was extracted from the figure of
		%the paper \cite{Trebbia2006}.}. 
		%The parameters  $[\mu/\hbar\omega_{\perp},k_{B}T/\hbar\omega_{\perp}]=$ 
		%(a) $[0.6, 2.3]$ (b) $[0.53, 3.0]$ (c) $[0.33, 3.3]$ (d) $[-6.78, 5.8]$.
    %Note the slight asymmetry in the experimental data of (c), which is not reproduced 
    %in the numerical data.
    Insets: quasi-1D SGPE contribution (Eq.\eqref{eq:q1dSGPE}; green dashed shaded region) 
		to the total linear density profiles (Eq.\eqref{eq:ntot}; black solid line).
		Bottom: Table showing parameters $\mu$ and $T$ for the quasi-1D SGPE 
  	density profiles which match the experimental data; 
		the parameters given in \cite{Trebbia2006} from a 3D Hartree-Fock 
	  fit are also shown.
    }
  \label{fig:HFF}
\end{figure*}%

To summarize briefly, our approach for modeling 
equilibrium properties of finite temperature
quasi-1D Bose gases is based on self-consistently solving Eq.\eqref{eq:q1dSGPE}
and Eq.\eqref{eq:nperp} 
for the desired total atom number (or measured peak density) and temperature
\footnote{As the SGPE solutions can be somewhat time-consuming, 
we find it convenient
(although by no means essential) to speed up the process by initially using the
modified Popov theory \cite{Andersen2002} 
(which matches the SGPE results very well \cite{AlKhawaja2002,Cockburn2011}) to 
arrive at a good initial condition for the self-consistent SGPE simulations.}.

%, as 
%in the prescription of [Kheruntsyan], each mode is treated as an individual 
%equilibrium 1D Bose gas, which therefore obeys Bose-Einstein statistics.
%Finally, as we solve the SGPE numerically, the lattice spacing 
%which arises in the discretized problem imposes an ultraviolet cutoff in energy. 
%We therefore choose a numerical grid, such that this energy cutoff is
%at $E_{\rm cut}=\hbar\omega_{\perp}$ (through the
%relation $E_{\rm cut}\approx2\pi\hbar\omega_z/(\Delta z)^2$),
%in order to be consistent with our scheme whereby we model only the 
%axial modes and transverse ground state with the SGPE.
%Additionally, the picture that atoms in transverse modes, with energies greater 
%than $\hbar\omega_{\perp}$, are treated as an equilibrium ideal Bose gas
%fits the conceptual picture of the SGPE,
%as atoms in these modes make up the heat bath to which the axial 
%subsystem is coupled, is also consistent with the SGPE formalism 
%we adopt here \cite{BijlsmaDuine2001}.
%---What about when k_{B} T < \hbar\omega_{\perp}?
In Sections \ref{sec:dprof} and \ref{sec:dfluct}, we
give a quantitative comparison between the proposed method
and published {\it in situ} experimental data, 
thereby highlighting the usefulness of this method.
For a direct comparison to these experiments, we choose here
to fix the temperature to that reported
in the experimental papers, and then use $\mu$ as a free parameter, which we 
vary until the required linear density is obtained at the trap centre.

\section{Comparison to Experiments}
\label{sec:comp}

\subsection{Density Profiles}
\label{sec:dprof}

We begin with a comparison of the model to total linear 
density profiles, as obtained by {\it in situ} absorption 
imaging within two experiments:
Sec. \ref{sec:dprof} gives a comparison to the data of Trebbia \etal \cite{Trebbia2006},
whose published analysis was based on a 3D Hartee-Fock model, %whereas Sec. \ref{sec:vanAmer}
before discussing the measurements of van Amerongen \etal \cite{vanAmerongen2008},
who instead analyzed their results via the modified Yang-Yang theory.

In each study, experimental data was compared to theory in order to 
quantify the equilibrium state through a 
chemical potential and temperature.
The temperature may be quite straight-forwardly measured by fitting 
the wings of the density distribution to the ideal gas result
(for sufficiently high temperatures), implying interactions
have little effect on measurements of this parameter.
Conversely, the chemical potential is far more dependent
on the model used in analyzing the 
%central region?) 
%of the 
density profile; this is because different theories represent
the full quantum Hamiltonian of the interacting system 
to different levels of approximation (see e.g. \cite{Proukakis2008}),
so incorporate many-body effects to a differing
degree.

It is worth pointing out also,
that the density profiles of the theory presented here,
and those from experimental
absorption imaging
share a common feature: namely, that 
additional analysis is required to identify
a phase coherent (or `true' condensate) 
and density coherent (or quasi-condensate)
fraction from the total density.
In the SGPE, the total density is due to both
coherent and incoherent particles,
however knowledge of the first and second order correlation functions 
was shown to be sufficient to isolate both quasi-condensate and
true-condensate densities \cite{Cockburn2011,Wright2011}.
In particular, the method discussed in \cite{Cockburn2011}
may be easily applied to
directly extract such components from {\em experimental} measurements
of correlation functions,
thereby offering a more accurate, experimentally self-consistent,
 characterisation of phase-fluctuating experiments (without the need to
resort to bimodal fits which become somewhat inaccurate in this limit).

\subsubsection{Comparison to work of Trebbia {\it et al.} \cite{Trebbia2006}}
\label{sec:Trebbia}

The results of Trebbia {\it et al.} demonstrated experimentally the breakdown 
of the Hartree-Fock method when applied to highly elongated Bose gases \cite{Trebbia2006}. 
It was concluded that this breakdown occurs because density fluctuations are not
accounted for accurately within this theory, since the energy lowering effect
of spatial correlations between atoms is not captured. These density correlations 
\cite{Naraschewski1999,Prokofev2002,Proukakis2006b,Bisset2009b}
are key to
correctly predicting the onset of quasi-condensation and the associated 
reduction in density
fluctuations; it was found, therefore, that the excited states did not saturate 
within Hartree-Fock theory and so no quasi-condensate was predicted to form.
\begin{figure*}[ht]
  \centering
  \begin{tabular}{c}
  \includegraphics[angle=0,scale=0.375,clip]{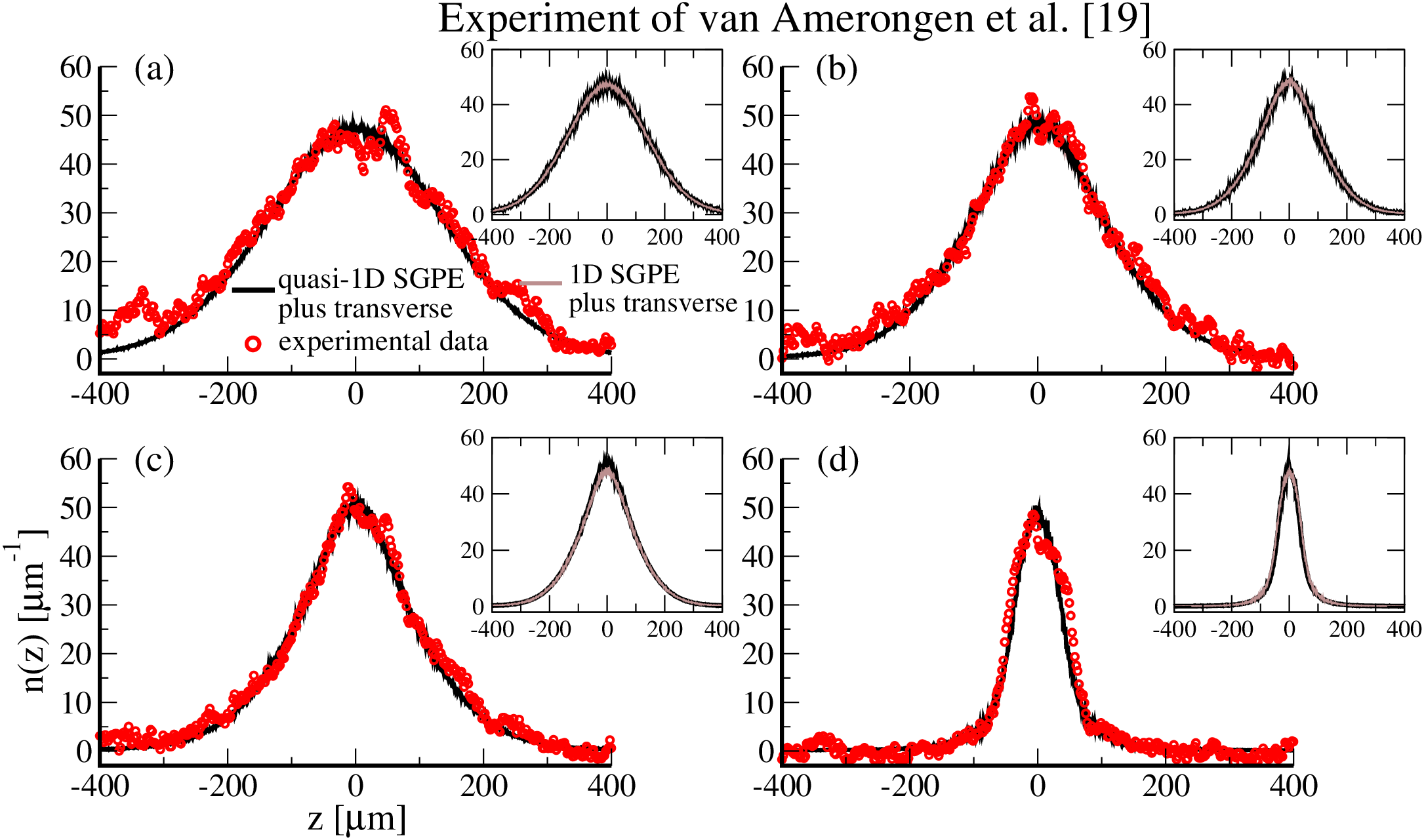}\\					
      \begin{tabular}{ c | c || c | c | c |}
      %Row 1
				 &  &
  	  	\multicolumn{3}{|c|}{$\mu/\hbar\omega_{\perp}$} \\
			\hline
      %Row 2
    		& \multirow{2}{*}{$k_{B}T/\hbar\omega_{\perp}$} & Analysis of \cite{vanAmerongen2008}  
   		  & \multicolumn{2}{|c|}{Current analysis} \\
        \cline{3-5}
      %Row 3
        &  & modified Yang-Yang & 1D SGPE & Quasi-1D SGPE \\
      \hline 
      \hline
      %Row 4 
      \multirow{1}{*}{Fig. $3$a} 
		  	& 4.0 & -2.74 & -2.74 & -2.74 \\%& -0.38 & 0 & 0.45 \\
	      \cline{2-5}
      \hline
      %Row 5
      \multirow{1}{*}{Fig. $3$b} 
		  	& 2.8 & -0.38 & -0.38 & -0.38 \\%& -0.38 & 0 & 0.45 \\
	      \cline{2-5}
      \hline
      %Row 6
      \multirow{1}{*}{Fig. $3$c} 
		  	& 2.4 & 0 & 0 & 0.03 \\%& -0.38 & 0 & 0.45 \\
	      \cline{2-5}
      \hline
      %Row 7
      \multirow{1}{*}{Fig. $3$d} 
		  	& 0.89 & 0.45 & 0.45 & 0.39 \\%& -0.38 & 0 & 0.45 \\
	      \cline{2-5}
      \hline
    \end{tabular}
  \label{table:YY_mus}

  \end{tabular}
  \caption[]{(Color online) 
	Top: Total linear density profiles from the quasi-1D SGPE (black solid line) 
  versus the experimental data of van Amerongen \etal \cite{vanAmerongen2008} (red circles).
  Insets: Quasi-1D SGPE (thick black line) versus 1D SGPE (thin brown line) 
  density profiles. 
  Bottom: Table showing parameters $\mu$ and $T$ for the 1D and quasi-1D SGPE 
  density profiles which match the experimental data and modified Yang-Yang model 
  fits from \cite{vanAmerongen2008}.} 
  \label{fig:YY}
\end{figure*}%

If we compare instead to the results of the quasi-1D SGPE, 
with the total linear density given by Eq.~\eqref{eq:ntot}, 
we see from Fig.\ref{fig:HFF} that the theoretical results (black solid line) 
match well those obtained within the experiments (red circles). 
The agreement is extremely good across the entire
range of temperatures considered, notably even at a temperature close to the 
crossover from quasi-condensate to thermal gas (Fig. \ref{fig:HFF}(c)). 
%\spcdid{Label figures with parameters and possibly SGPE fraction OR qc fraction OR both.}
It is precisely this regime of critical fluctuations 
%\spcdid{Ginzburg criterion?}
in which a deviation from mean field theory might be expected, due to the 
lack of a well defined mean field quantity. 
%In fact, upon comparing the stochastic
%results to the modified Popov theory 
%\spcdid{to include in figure?}
%of \cite{Andersen2002}, we find good agreement with the SGPE results
%in all cases with the exception of Fig.\ref{fig:HFF}(c),
%where the SGPE density peak is slightly higher in the centre
%and closer to the experimental result.
Interestingly, behaviour suggestive of this was found 
in \cite{Trebbia2006} 
when comparing their data to the Hartree-Fock mean field model: 
the experimental data was found to have a higher peak than 
the mean field result [see Fig1(c) of \cite{Trebbia2006}],
which illustrates the potential importance of 
including many-body effects,
as studied recently in the context of a
finite-temperature classical field theory \cite{Wright2011}.

%The Table shows the chemical potentials we find give 
We obtain a good 
fit between the quasi-1D SGPE and experimental density profiles, 
at the experimentally measured temperatures, 
with a comparison between the chemical potentials extracted in our treatment
and the published values based on the 
Hartree-Fock analysis of \cite{Trebbia2006} shown in the table
\footnote{Note that in comparing the SGPE data to
that of \cite{Trebbia2006}, we have: 
(i) shifted the SGPE density profiles by $8\mu$m to the right
in order to match the position of the experimental peak densities;
(ii) used the relation $\mu=\mu_{\rm 3D}-\hbar\omega_{\perp}$, with 
$\mu_{\rm 3D}$ the chemical potential reported in \cite{Trebbia2006}.}.
The deviation between the parameters of different theoretical methods
should not be of any concern, as it merely
highlights $\mu$ as a model dependent quantity, 
i.e. it is dependent on the actual Hamiltonian used to analyse the experimental results,
and therefore
varies depending upon the level of 
approximation \cite{Wright2011,Cockburn2011}.

The insets of Fig.\ref{fig:HFF} show the contributions to the total
linear density profiles due to the axial SGPE density (green dashed, shaded region),
with the remainder coming from Eq.\eqref{eq:nperp}. 
The importance of the SGPE contribution is clear 
in the first three plots, which shows an appreciable
fraction of atoms reside in axial modes for these parameters.
In the highest temperature case, 
shown in Fig.\ref{fig:HFF}(d), 
$\mu<0$ and the gas is entirely in the thermal phase;
here the density is instead due almost entirely to 
atoms in tranverse modes.

\subsubsection{Comparison to work of van Amerongen {\it et al.} \cite{vanAmerongen2008}}
\label{sec:vanAmer}

The second experiment that we consider is that of van Amerongen {\it et al.} 
\cite{vanAmerongen2008}.
This was the first experimental comparison to the exact Yang-Yang thermodynamic
solution to the finite temperature 1D Bose gas problem \cite{Yang1969},
also referred to as the thermodynamic Bethe ansatz.

In \cite{vanAmerongen2008}, the one-dimensional Yang-Yang theory was used to 
represent the axial modes
and transverse ground state (of width $l_{\perp}$), 
while the contribution to the linear density due
to atoms in transverse excited states, was accounted for using the method 
we also adopt here. The total density profiles in \cite{vanAmerongen2008} 
were therefore calculated using Eq.\eqref{eq:ntot}, 
with the role of the SGPE contribution $\langle|\psi|^{2}\rangle$ 
instead played by the 1D Yang-Yang prediction.
Therefore, in comparing to this work, we will gain insight on two fronts:
firstly, how well the SGPE matches the experimental data in this regime,
and simultaneously (but indirectly), how well the SGPE prediction for the density profiles 
matches that due to Yang-Yang thermodynamics.

Fig.\ref{fig:YY} shows that the agreement between the 
experimental data and the proposed quasi-1D SGPE approach is again very good across the entire
temperature range probed, including the crossover from quasi-condensate 
to degenerate thermal gas. 
%\spcdid{Label figures with parameters and possibly SGPE fraction OR qc fraction OR both}. 
%From Fig.\ref{fig:one} this might be expected, since we are probing smaller
%values of $\mu$ and $k_{B}T$, relative to the transverse
%energy $\hbar\omega_{\perp}$, than in matching the data
%of Trebbia {\it et al.} in Section \ref{Trebbia}. 

%For completeness, in the insets of Fig.\ref{fig:YY}, we compare
We now wish to discuss how our quasi-1D SGPE results compare to those from the SGPE
with the usual 1D equation of state. 
Practically, this means using the equilibrium result 
of Eq.~\eqref{eq:1dSGPE}, rather than Eq.~\eqref{eq:q1dSGPE}, as the axial density 
input $\langle|\psi|^{2}\rangle$ in Eq.~\eqref{eq:ntot}. 
Although this approach also recovers closely the total density profiles
found with the quasi-1D SGPE at each temperature (insets to Fig. \ref{fig:YY}),
and therefore also those measured experimentally, 
it is important to note that 
%they 
each approach can 
lead to slightly different chemical potentials 
for the same temperature.

%The parameters for each case are highlighted in the table 
%of Fig.\ref{fig:YY}.
%We see that the 1D and quasi-1D two models predict the gas to 
%be in slightly different thermodynamic states, 
%i.e. $n(z;\mu,T)$ is not the same in the 
%1D and quasi-1D SGPE models, 
%since different values of $\mu$ are required if 
%we are to match the experimental density profile at each experimental $T$.
%In Section \ref{sec:dfluct}, this distinction will become important in
%predicting density fluctuations consistent with the experimental findings
%of \cite{Armijo2010}, and ultimately, we find that this observable proves 
%to be a good measure for testing which model is most consistent.

Importantly, we find that the parameters used 
to obtain the 1D SGPE results are {\em identical} to those 
obtained from fits of the modified Yang-Yang model to the density data in \cite{vanAmerongen2008}.
These results have also been reported 
\cite{Kheruntsyan2010} to arise 
within the context of the closely-related 1D stochastic 
projected Gross-Pitaevskii equation (SPGPE)
\cite{Blakie2008} in parallel independent work,
which also looked at the momentum distribution of the gas
after focussing \cite{Shvarchuck2002}. 
This provides an 
indirect additional test between the 1D SGPE, 
the 1D SPGPE 
%\spcdid{I would not claim this necessarily,
%as we give no direct evidence here}
and Yang-Yang theories in the weakly-interacting regime.
The parameters predicted by the quasi-1D SGPE, 1D SGPE and modified Yang-Yang
models are shown in the Table of Fig.\ref{fig:YY}.

Having established that both the 1D and quasi-1D SGPE approaches
accurately reproduce experimental density profiles
(and therefore also those due to the modified Yang-Yang approach used in \cite{vanAmerongen2008}),
we now turn to an investigation of density fluctuations,
which provide a more sensitive probe for the validity of these theories.

\subsection{Density fluctuations}
\label{sec:dfluct}

%Density fluctuations represent a quantity dependent upon
%higher order operator moments than the average density; 
%such higher order moments 
%become important when dealing with highly elongated Bose gases, as they 
%are crucial in correctly predicting the onset of the quasi-condensation,
%as was found in \cite{Trebbia2006}. 
%Having established that both the 1D and quasi-1D SGPE approaches
%are able to reproduce experimentally obtained density profiles,
%and therefore also those due to the MYYM approach used in \cite{vanAmerongen2008},
%we now turn to recently obtained data measuring density fluctuations within 
%finite temperature Bose gases, as a further test of these methods.

%Within an ideal Bose gas picture,
Density fluctuations are increased markedly within an ideal Bose gas,
due to an effect of quantum statistics, which leads to 
atomic bunching \cite{LandauLifshitz_statmech1}.
However, at sufficiently low temperatures, and in the presence of interactions,
quasi-condensation leads to atom-atom 
correlations which overcome the tendency for bosonic atoms to bunch together, 
and therefore to a reduction in the level of density fluctuations, relative to 
those expected in an ideal Bose gas \cite{Esteve2006}.

\subsubsection*{Comparison to work of Armijo {\it et al.} \cite{Armijo2010}}

In a recent paper \cite{Armijo2010}, Armijo {\it et al.} measured the second and 
third moments of the density fluctuations of a finite temperature Bose gas,
comparing these to theoretical predictions from ideal Bose gas
and quasi-condensate mean-field models, 
and also the modified Yang-Yang model of \cite{vanAmerongen2008}. 
We now briefly outline their experimental method 
before describing the 
numerical scheme we follow in order to
closely mimic this.
%manner in which we 
%closely follow these in carrying out the SGPE simulations.
%Within an ideal Bose gas picture,
%density fluctuations are increased markedly within a Bose gas
%due to the effect of quantum statistics, which leads to the
%phenomena of atomic bunching [LandauLifshitzPt1].
%At sufficiently low temperatures, however, quasi-condensation leads to atom-atom 
%correlations which overcome the tendency for bosonic atoms to bunch together, 
%and therefore to a reduction in the level of density fluctuations, relative to 
%those expected in an ideal Bose gas \cite{Esteve2006}.

\emph{Experimental procedure:} A harmonic trapping potential 
leads to a density profile in which the number of atoms varies spatially, 
and therefore close to the trap centre it is possible to have a scenario
where there is sufficient levels of degeneracy such that a quasi-condensate is
formed, whereas in the low-density wings, the gas is still 
effectively a non-interacting thermal gas.
%Owing to this, it is possible to examine each of the regimes
%by scanning the spatial extent of the trapped gas. 
Thus, at a single temperature, by scanning the spatial extent of the trapped gas,
it is possible to observe both the 
enhancement of density fluctuations, due to 
quantum statistics (low density, ideal Bose gas), 
and their subsequent suppression, due to particle interactions 
(higher density, quasi-condensate regime).

An approach based on this observation 
was first undertaken experimentally in 
\cite{Esteve2006}, and subsequently followed 
by more detailed studies in \cite{Armijo2010,Armijo2011}.
In \cite{Armijo2010}, the gas was probed using a CCD camera, which 
effectively divides observations of the gas into
pixel sized regions (of size $\Delta=4.5\mu m $ in this case). 
Absorption imaging allowed for the number of atoms within each pixel, 
$ N $, to be measured, and %by carrying this out many times,
repeated measurements provided a set of fluctuating values,
as well as an average
% were averaged over to yield an average 
number per pixel, $\langle N \rangle$. 
The $p$-th moment of the density fluctuations 
for the set of measurements were then 
calculated for each pixel as usual 
via 
$  \langle \delta N ^{p}\rangle=\left\langle \left( N - \langle N \rangle \right)^p \right\rangle$.

The aim of the present work is to demonstrate the SGPE as an {\it ab initio} 
method for analyzing experimental findings, 
and so we wish to implement a numerical scheme 
which follows experimental procedures as closely as possible.
Fortunately, the grand canonical formulation of the SGPE
makes it relatively simple to simulate the experimental 
methods used in \cite{Armijo2010}.
This is because, in addition to the 
unified treatment of both (quasi-)condensate 
and thermal atoms in density profiles, the SGPE shares a
second feature in common with experiments, namely
a shot-to-shot variation between individual realizations. 
This variation enables us to straightforwardly model the equilibrium density
fluctuation experiments of \cite{Armijo2010},
but is also important in dynamical studies
\cite{Cockburn2010,Weiler2008,Damski2010,Das2011}.
%, as we now discuss.
%and we mimic their method numerically using the SGPE single run data 
%in lieu of individual experimental runs. 

\emph{Numerical SGPE procedure:}
Physical observables within the SGPE are obtained as products
of the stochastic wavefunction $\psi$, averaged over many realizations of
the noise $\eta$ (see Eq.~\ref{eq:q1dSGPE}, and Ref.~\cite{Cockburn2009} for a simple overview).
This leads naturally to a shot-to-shot variation
between numerical realizations,
in a way analogous to an individual experimental run.
%; each realization may be viewed as a member of
%a grand canonical ensemble.
%This can be viewed as a grand canonical ensemble made up 
%of many different realizations of the same system.
%The average density, for example, 
%$n=\langle\psi^{\dagger}\psi\rangle$, where $\langle\ldots\rangle$
%denotes a sum over 
%
%To model all aspects of the experiment here, we undertake the exact same 
%study numerically, wherein 
By treating each noise realization like an experimental realization,
%To model each SGPE numerical run mimics 
we are able to accurately synthesize the experimental procedure 
of \cite{Armijo2010} within our numerical simulations. 
We emphasise, however, that within this
analogy 
% between single numerical runs and single experimental realizations
it should be understood that single runs
represent weighted contributions to
%only make sense when contributing to
%in the sense that 
% numerical moments like $ \langle \delta N ^{p}\rangle $ should 
averages over fluctuating quantities, %such as moments of the density fluctuations, 
and that it is only such appropriately 
averaged quantities that 
we expect to match well those determined experimentally, 
as indeed is found to be the case below.

\begin{figure}[bt!]
  \centerline{
    \includegraphics[angle=0,scale=0.255,clip]{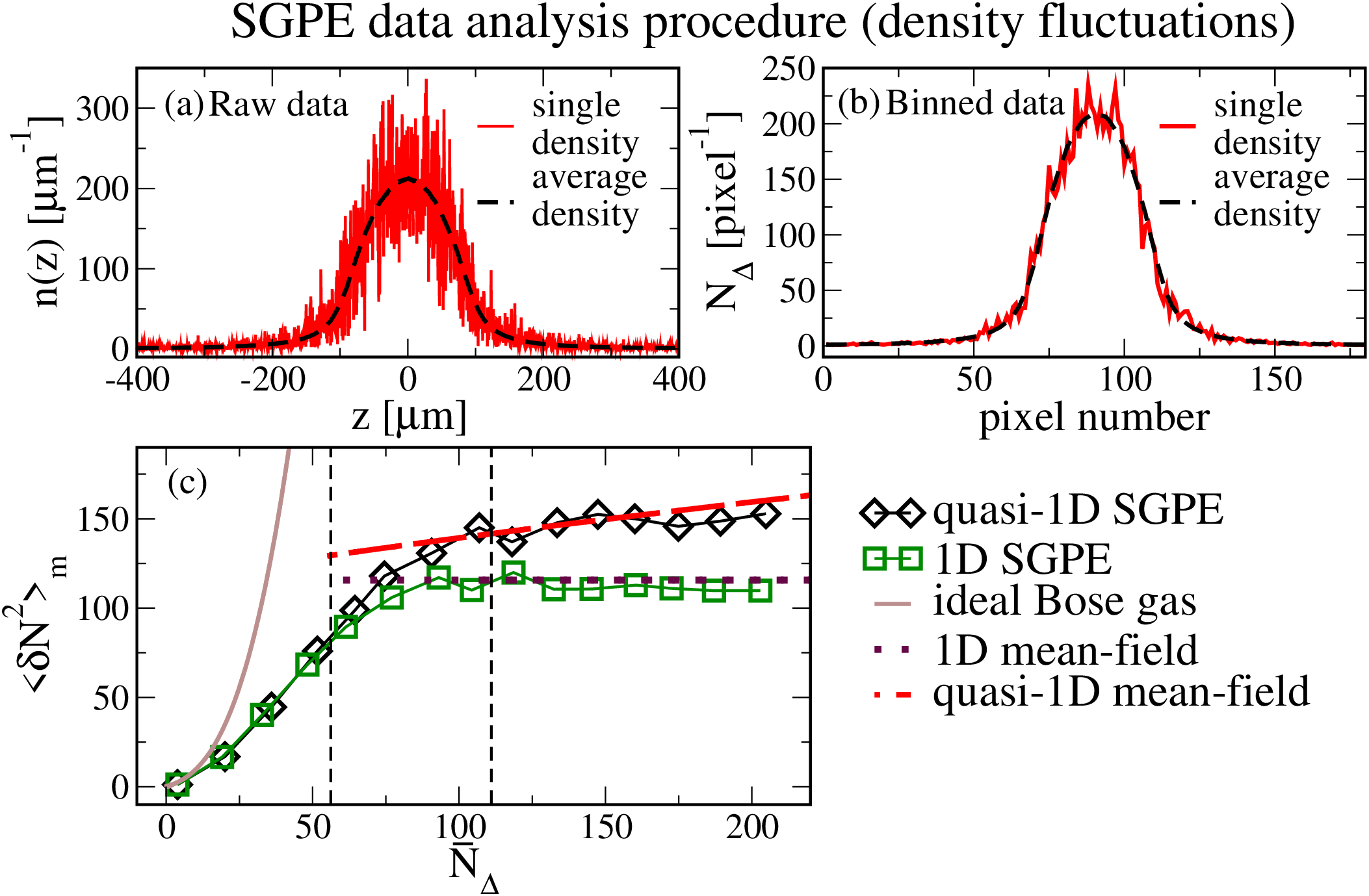}
  }
  \caption[]
    {
    (Color online) 
    Top row: quasi-1D SGPE density (noisy red curve)
    obtained from (a) a single numerical run (raw data),
    and (b) corresponding spatially binned data; dashed black
    curves display the density averaged over 1000 independent realisations.
    (c) $\langle \delta N ^{2}\rangle_{m}$ \cite{endnote86} from the binned quasi-1D
    SGPE data (black diamonds) and 1D SGPE data (green squares), compared against
    correponding mean field results for an ideal Bose gas (brown solid line)
		1D quasi-condensate (dotted maroon, horizontal) and quasi-1D quasi-condensate 
    (dashed red); the thin, vertical dashed lines indicate the `crossover' region
    where the interaction and thermal energies become comparable. This data is for
		$T=96$nK. 
    }
  \label{fig:method}
\end{figure}

Our numerical procedure is to
run a large number of stochastic simulations ($1000$), 
each of which yields a fluctuating density profile. 
We measure the number fluctuations within each pixel 
sized region, by first spatially binning 
the SGPE density data 
into $\Delta$-sized regions \footnote{
The spatial numerical grid spacing, $\Delta z$, is chosen to give the 
desired ultraviolet energy cutoff ($=\hbar\omega_{\perp}$), 
which is much smaller
than the pixel size}, 
in order to give an output consistent 
with that obtained in the experiments (see Fig.\ref{fig:method} (a)-(b)). 
Integrating over the numerical grid points 
within a single pixel yields $N_{z,\Delta}$,
the value for the (fluctuating) atom number from a single stochastic realization, within that pixel. Repeating the same procedure for 
the mean total density generates a binned 
average \emph{axial} pixel number, $\bar{N}_{z,\Delta}$. 
The contribution from the axial modes to the second 
($p=2$) and third ($p=3$) moments of the density fluctuations are 
then calculated as 
%\begin{equation}
$  \langle \delta N ^{p} \rangle_{z}=\left\langle \left(N_{z,\Delta}-\bar{N}_{z,\Delta}\right)^p \right\rangle$.

Similarly, a binned transverse contribution to the average 
pixel number $\bar{N}_{\perp,\Delta}$ may be obtained, 
and the \emph{total} average atom number 
in each pixel is then given by 
$\bar{N}_{\Delta}=\bar{N}_{z,\Delta}+\bar{N}_{\perp,\Delta}$ 
(which corresponds to $\langle N \rangle$ in \cite{Armijo2010}).
%
%\end{equation}
%where $p=2$ and $p=3$ for the second and third moments, respectively.
%to mimic all stages of the experimental data analysis, we undertake the following two steps in analysing
%our data:
%firstly, we compute the total density fluctuations 
%\begin{equation}
%$  \langle \delta N ^{p}\rangle=\langle \delta N ^{p}\rangle_{z}+\langle N \rangle_{\perp} $,
%\end{equation}
%where $\langle N \rangle_{\perp} = ???$.
%Moreover, we account for the finite spatial resolution of the 
%experiment (or our numerical binning ???) via the relation
%\begin{equation}
%$  \langle\delta N ^{p}\rangle_{m} = \kappa_{p} \langle \delta N ^{p} \rangle$,
%\end{equation}
%where $\kappa_p = ...$.
%{\em A BIT LOST AFTER HERE - UNTIL CONCLUSIONS!}
As we treat the atoms in the transverse modes in a static way, they give a
non-zero contribution only to average properties, and so do not 
contribute to moments of the density fluctuations directly. 
However, as we found in Section \ref{sec:dprof},
atoms in these modes are well approximated as a degenerate ideal gas, for which
$\langle \delta N ^{2}\rangle_{\perp} \simeq \langle \delta 
N ^{3}\rangle_{\perp} \simeq \langle N \rangle_{\perp}$ \cite{Armijo2010},
and we therefore assume that these atoms contribute a factor 
$\langle N \rangle_{\perp}$ to both second and third moments. 
So, ultimately, we compute the total density fluctuations as
%\begin{equation}
$\langle \delta N ^{p}\rangle=\langle \delta N ^{p}\rangle_{z}+\langle N \rangle_{\perp}$,
%\label{dN2}
%\end{equation}
for $p=2,3$. 

To illustrate our procedure, we plot in Fig.\ref{fig:method} 
an example single run and average density
profile both before (Fig.\ref{fig:method}(a)), and after 
(Fig.\ref{fig:method}(b)), spatial binning. 
Comparing these plots, it is clear that the binning 
procedure significantly smooths 
the raw single run data, as would be 
expected for this kind of a spatial 
coarse graining procedure;
note that the binned data looks remarkably similar to the
experimental fluctuating density profile shown in \cite{Armijo2010}
(Fig. 1(c) of that work).
We additionally
show in Fig.\ref{fig:method}(c) the
variance in the density fluctuations which results
from 
%Eq.\eqref{dN2}, applied to 
the set of $1000$ 
fluctuating binned densities; this is plotted
against the average number of atoms per pixel 
\footnote{As detailed in \cite{Armijo2010}, we note there is a factor which 
relates the experimentally measured moments, 
$\langle\delta N\rangle_{m}$ to those obtained theoretically, via
%\begin{equation}
$\langle\delta N ^{p}\rangle_{m} = \kappa_{p} \langle \delta N ^{p} \rangle$.
%\end{equation}
The factor $\kappa_{p}$ arises due to the finite spatial resolution of the 
experiment, and therefore we must scale our findings 
in order to account for this experimental issue.\label{endnote137}}.

Making use of the thermodynamic relation 
$\langle\delta N ^2\rangle=k_{B}T\Delta(\partial n /\partial \mu)_T$ 
\cite{LandauLifshitz_statmech1},
it is possible to derive mean field results for the
density fluctuations,
%valid in both %high and low density regimes,
based on both the ideal gas and quasi-condensate equations of state.
The ideal gas result is \cite{Armijo2011}
\begin{equation}
  \langle\delta N ^2\rangle=\frac{1}{\lambda_{\mathrm{dB}}}
  \sum_{j=1}^{\infty}\frac{\sqrt{j}}{k_{B}T}\frac{e^{j\mu/k_{B}T}}{\sqrt{j}}
  \frac{1}{\left(1-e^{-j\hbar\omega_{\perp}/k_{B}T}\right)},
\end{equation}
while using the 1D equation of state 
for a quasi-condensate ($\mu[n]=g n$)
gives the simple result \cite{Esteve2006}
%\begin{equation}
$\langle\delta N ^2\rangle_{\rm 1D}={k_{B}T\Delta}/{g}$.
%\end{equation}
Using instead the quasi-1D equation of state, Eq.\eqref{eq:q1d_eos}, yields,
%\begin{equation}
$\langle\delta N ^2\rangle_{\rm quasi-1D}=\langle\delta N ^2\rangle_{\rm 1D}\left[1+{(\mu-V(z))}/{\hbar\omega_{\perp}}\right]$.
  %\frac{k_{B}T\Delta}{g}\left[\frac{\mu-V(z)}{\hbar\omega_{\perp}}+1\right].
%\end{equation}
%We see from Fig.[] that the quasi-1D SGPE data matches well the shot noise result 
%at very small $N$, before crossing over to the ideal Bose gas prediction,
%due to the effects of atomic bunching.

Within the higher density region of Fig.\ref{fig:method}(c), 
we show the mean field results due to the
1D and quasi-1D equations of state for a quasi-condensate.
It is clear that the 1D SGPE shows good agreement with the 1D mean field
result, whereas the quasi-1D SGPE instead agrees very well 
with the quasi-1D mean field prediction.
The two vertical lines indicate the region
%$\Delta k_{B}T/ 4 g < \langle N \rangle < \Delta k_{B}T/ 2 g$,
where %the boundaries correspond to the densities
the interaction energy 
%thermal particle ($=2gn$)
%and a quasi-condensate particle ($=gn$) 
and the average thermal energy become comparable.
For higher densities, 
interactions significantly reduce 
$\langle\delta N ^2\rangle$ below the ideal 
gas prediction, and we see this is
more pronounced in the 1D SGPE case
relative to the quasi-1D SGPE data,
as expected from the 1D and quasi-1D mean field
predictions.
\begin{figure}[t!]
  \centerline{
    \includegraphics[angle=0,scale=0.29,clip]{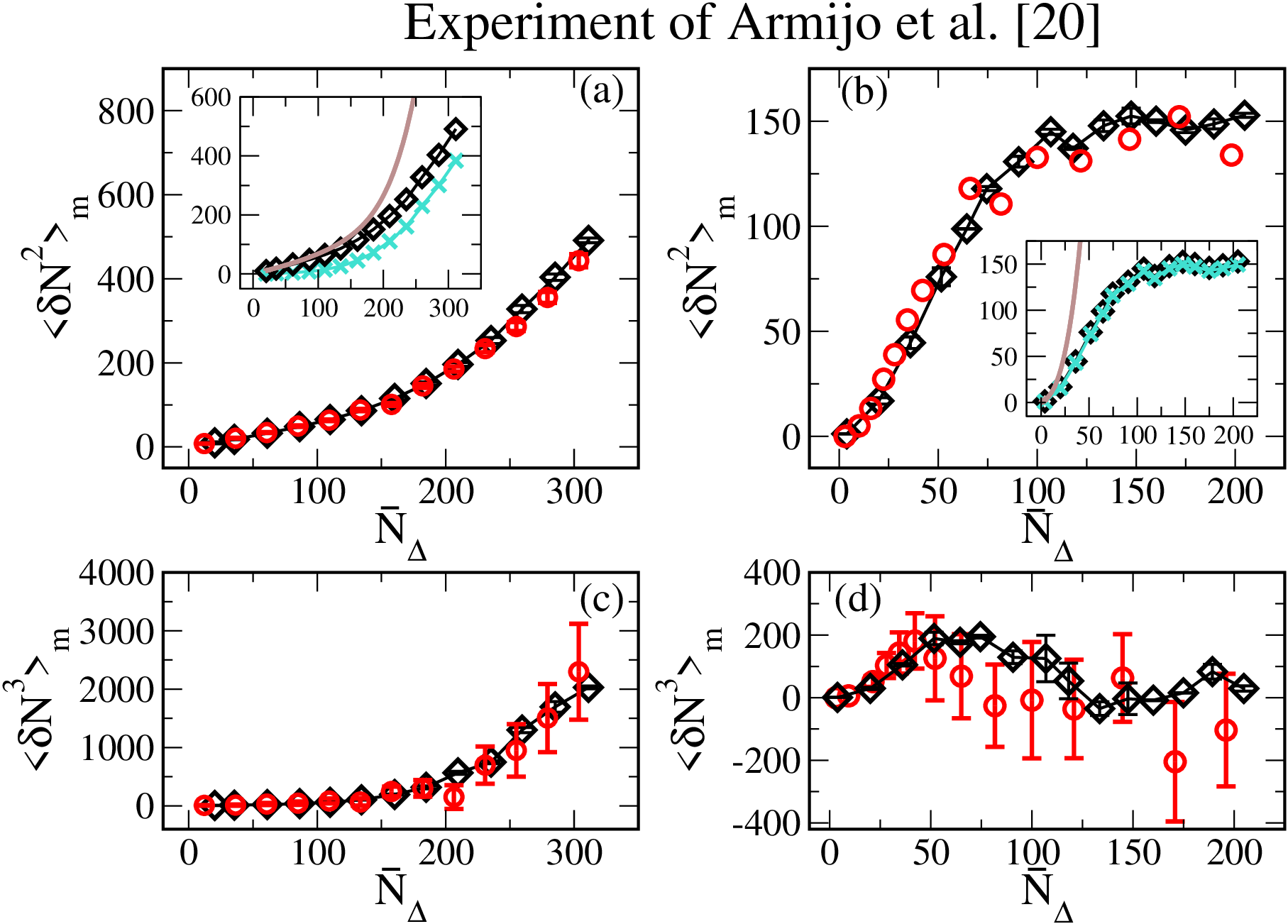}
  }
  \caption[]
    {
    (Color online) 
		Second (top row: (a)-(b)) and third (bottom row: (c)-(d)) moments of the atom number 
    fluctuations from the quasi-1D SGPE data (black diamonds)
    %the 1D SGPE (green squares) 
    and experimental
    data from the paper of Armijo \etal \cite{Armijo2010} (red circles). 
		%\footnote{
		%The data of Figs. \ref{fig:Armijo}(a) and (c) were supplied by Isabelle 
		%Bouchoule, while for Figs. \ref{fig:Armijo}(b) and (d) the data were
		%extracted from the paper. 		
		%}). 
    Temperatures are $T=376$nK (left images) and $T=96$nK (right images). 
    Insets: axial (light blue crosses) versus axial
    plus transverse (black diamonds) contributions to the
    number fluctuations vs. the
    ideal gas result (brown solid line). 
    %Also shown are mean field results for the ideal Bose gas
    %(dot-dashed blue) and the quasi-condensate (1D, dotted maroon; 
    %quasi-1D, dot-dashed brown).
    }
  \label{fig:Armijo}
\end{figure}

A key point from our analysis at this temperature, 
is that while the ideal gas
equation of state is valid only for small densities, 
and the mean-field quasi-1D equation of state
holds at high densities, 
the quasi-1D SGPE, like the experimental data, 
provides a smooth crossover between each of these regimes. 
This is because, in moving outwards from the centre of the trap in the presence of 
a quasi-condensate, at some point the gas changes phase to a thermal gas, 
and a mean field theory cannot be expected to accurately describe 
fluctuations in the transition region 
near the edge of the quasi-condensate.

\emph{Comparison to Experiment:}
Fig.\ref{fig:Armijo} shows a comparison of the experimental 
density fluctuation data obtained by Armijo {\it et al.} (red circles), 
and the quasi-1D SGPE model (black diamonds). 
We plot in the top row $\langle\delta N ^{2}\rangle_{m}$,
and in the bottom row $\langle\delta N ^{3}\rangle_{m}$, 
each versus $\bar{N}_{\Delta}$ for two temperatures 
($T=376$nK (left) and $96$nK (right)).

Concentrating on the $\langle\delta N ^{2}\rangle_{m}$ data first,
%(Fig.\ref{fig:Armijo}, right) we see a number of 
%interesting features: firstly, 
it is clear that the quasi-1D SGPE numerical results follow the
experimental data well in both the high temperature (left) and low
temperature cases (right images).
%$\bar{N}_{\Delta}$ ($\lesssim 50$), which corresponds to regions in 
%the wings of the density distributions. 
Notice that as the high-density, quasi-condensate regime is reached
in Fig.\ref{fig:Armijo}(b), 
the 1D SGPE result (shown in Fig.\ref{fig:method}) would predict
too great a reduction in density 
fluctuations relative 
to the experimental results, 
whereas the quasi-1D SGPE 
captures the experimental behaviour very well. 
Physically, this suggests that the effect of the transverse swelling 
of the quasi-condensate near the centre of the trap cannot be ignored
for these parameters, and that the quasi-1D extension to the SGPE
is therefore essential here.

Armijo {\it et al.} found a similar trend when comparing between their
experimental data and the modified Yang-Yang model.
While the Yang-Yang result matched the low density data well, 
at higher densities it displayed the same behaviour as
the 1D SGPE result.
Instead, Armijo {\it et al.} found the quasi-1D 
mean field result to better reproduce 
the experimental behaviour in the 
higher density quasi-condensate regime.
So, as for the density profiles, 
we again find the 1D SGPE data 
to reproduce the 1D Yang-Yang behaviour, whereas the quasi-1D SGPE
goes beyond the modified Yang-Yang model by capturing the quasi-1D behaviour found in the experiment
in all regimes probed.

The results for the third moment (Fig.\ref{fig:Armijo}, bottom row), 
$\langle\delta N ^{3}\rangle_{m}$, also show good agreement between the
quasi-1D SGPE and the experimental data (within the large experimental error bars).
%In contrast to the $\langle\delta N ^{2}\rangle_{m}$ data, 
%show there is little to distinguish
%between the 1D and quasi-1D models for this quantity, at this temperature.
%In both the 1D and quasi-1D cases, however, the SGPE data shows good 
%agreement with the experimental results, as for the density profiles 
%of Section \ref{sec:dprof}.

%For the higher temperature data (Fig.\ref{fig:Armijo}, left), 
%the quasi-condensate makes up a far smaller fraction of 
%the system, which explains the comparatively smaller 
%reduction in density fluctuations relative to the ideal 
%gas prediction (brown dashed line),
%compared to that found in the low 
%temperature case.
%Because of this behaviour, the high temperature data 
%
%PLEASE CHECK:
The insets to Figs. \ref{fig:Armijo}(a)-(b) show clearly that the reduction
in density fluctuations compared to the ideal gas prediction (brown
solid line)
is more pronounced in the low temperature case, where there is a higher
quasi-condensate fraction. Contrary to this, the enhanced presence of thermal atoms
in the high temperature case,
provides an opportunity to test whether our approximate treatment of
the transverse mode contribution to $\langle\delta N ^{2}\rangle_{m}$, 
leads to the correct behaviour at low density.
The results before (light blue crosses) and after (black diamonds) 
the transverse contribution is added
are shown in the insets to Figs.\ref{fig:Armijo}(a)-(b). 
This addition results in a noticeable shift upwards 
in the data of Fig.\ref{fig:Armijo}(a)
while for the lower temperature
case, the difference following this addition is negligible.
Comparing the SGPE data to the expected low density result, 
the ideal gas prediction for the same parameters 
(solid brown line), 
%which corresponds to the shot noise level $\langle N \rangle$ associated 
%to the {\it total} density
we see that adding a contribution $\langle N \rangle_{\perp}$ to 
the axial contribution fully captures 
the expected behaviour.
%whereas for the low temperature data of Figs.\ref{fig:Armijo}(b,d)
%there is only a negligible contribution due to the transverse excited modes,
%the contribution from atoms in these modes for the higher temperature case
%is more appreciable.

\section{Conclusions}

In conclusion, we have proposed and implemented
a suitably modified 
one-dimensional stochastic Gross-Pitaevskii equation, 
which was shown to provide excellent {\it ab initio} predictions
for both {\it in situ} experimental density profiles obtained by Trebbia \etal 
\cite{Trebbia2006} and van Amerongen \etal \cite{vanAmerongen2008},
and {\it in situ} density fluctuation data from the experiment 
of Armijo \etal \cite{Armijo2010}.
This was achieved by 
matching peak densities (equivalent to total atom number)
to a hybrid scheme, which combines the aforementioned stochastic
model
with a previously reported approach based on 
treating transverse thermal modes
as independent ideal Bose gases.

The study of density fluctuations showed that our combined approach
captures all experimental regimes studied in a unified manner, smoothly
interpolating between mean field models, whose individual validity is
restricted to either the low density or high density regimes.
Importantly, it was found that analyzing individual stochastic 
realizations in the same way as individual experimental runs, led to
good agreement between the density statistics in each case.

Reducing our stochastic model to the previously tested one-dimensional stochastic
Gross-Pitaevskii equation showed that:
(i) the latter model is consistent with Yang-Yang predictions (in the weakly-interacting
regime probed here), and that
(ii) while both one-dimensional and quasi-one-dimensional approaches accurately 
reproduce equilibrium density profiles, they do so with slightly different chemical 
potentials.
%(which relates to the dimensionality, rather than the model employed).

The confidence gained from these successful comparisons suggests that the quasi-one-dimensional
stochastic Gross-Pitaevskii model, which was proposed here as a hybrid
of different previously implemented approaches, is an excellent model for describing quasi-condensate
experiments at equilibrium; the extent to which this model can be directly applied to study
dynamical features, such as experimentally-measured properties after expansion, 
remains to be investigated.

\bigskip
\paragraph*{Acknowledgments.}
We would like to thank Isabelle Bouchoule, and Aaldert van Amerongen 
and Klaasjan van Druten, for providing their experimental data for use 
in our comparison.
NPP acknowledges discussions with Dimitri Frantzeskakis on the Gross-Pitaevskii equation
in the quasi-one-dimensional regime, and Matt Davis, Geoff Lee and Karen Kheruntsyan
for discussions on their related theoretical modeling. 
We also thank Carsten Henkel and Antonio Negretti
for related discussions. This project was funded by the EPSRC.

\bibliographystyle{apsrev}
\bibliography{twa-sgpe}

\begin{thebibliography}{84}
\expandafter\ifx\csname natexlab\endcsname\relax\def\natexlab#1{#1}\fi
\expandafter\ifx\csname bibnamefont\endcsname\relax
  \def\bibnamefont#1{#1}\fi
\expandafter\ifx\csname bibfnamefont\endcsname\relax
  \def\bibfnamefont#1{#1}\fi
\expandafter\ifx\csname citenamefont\endcsname\relax
  \def\citenamefont#1{#1}\fi
\expandafter\ifx\csname url\endcsname\relax
  \def\url#1{\texttt{#1}}\fi
\expandafter\ifx\csname urlprefix\endcsname\relax\def\urlprefix{URL }\fi
\providecommand{\bibinfo}[2]{#2}
\providecommand{\eprint}[2][]{\url{#2}}

\bibitem[{\citenamefont{Bloch et~al.}(2008)\citenamefont{Bloch, Dalibard, and
  Zwerger}}]{Bloch2008}
\bibinfo{author}{\bibfnamefont{I.}~\bibnamefont{Bloch}},
  \bibinfo{author}{\bibfnamefont{J.}~\bibnamefont{Dalibard}}, \bibnamefont{and}
  \bibinfo{author}{\bibfnamefont{W.}~\bibnamefont{Zwerger}},
  \bibinfo{journal}{Rev. Mod. Phys.} \textbf{\bibinfo{volume}{80}},
  \bibinfo{pages}{885} (\bibinfo{year}{2008}).

\bibitem[{\citenamefont{G\"orlitz et~al.}(2001)\citenamefont{G\"orlitz, Vogels,
  Leanhardt, Raman, Gustavson, Abo-Shaeer, Chikkatur, Gupta, Inouye, Rosenband
  et~al.}}]{Gorlitz2001}
\bibinfo{author}{\bibfnamefont{A.}~\bibnamefont{G\"orlitz}},
  \bibinfo{author}{\bibfnamefont{J.~M.} \bibnamefont{Vogels}},
  \bibinfo{author}{\bibfnamefont{A.~E.} \bibnamefont{Leanhardt}},
  \bibinfo{author}{\bibfnamefont{C.}~\bibnamefont{Raman}},
  \bibinfo{author}{\bibfnamefont{T.~L.} \bibnamefont{Gustavson}},
  \bibinfo{author}{\bibfnamefont{J.~R.} \bibnamefont{Abo-Shaeer}},
  \bibinfo{author}{\bibfnamefont{A.~P.} \bibnamefont{Chikkatur}},
  \bibinfo{author}{\bibfnamefont{S.}~\bibnamefont{Gupta}},
  \bibinfo{author}{\bibfnamefont{S.}~\bibnamefont{Inouye}},
  \bibinfo{author}{\bibfnamefont{T.}~\bibnamefont{Rosenband}},
  \bibnamefont{et~al.}, \bibinfo{journal}{Phys. Rev. Lett.}
  \textbf{\bibinfo{volume}{87}}, \bibinfo{pages}{130402}
  (\bibinfo{year}{2001}).

\bibitem[{\citenamefont{Rychtarik et~al.}(2004)\citenamefont{Rychtarik,
  Engeser, N\"agerl, and Grimm}}]{Rychtarik2004}
\bibinfo{author}{\bibfnamefont{D.}~\bibnamefont{Rychtarik}},
  \bibinfo{author}{\bibfnamefont{B.}~\bibnamefont{Engeser}},
  \bibinfo{author}{\bibfnamefont{H.-C.} \bibnamefont{N\"agerl}},
  \bibnamefont{and} \bibinfo{author}{\bibfnamefont{R.}~\bibnamefont{Grimm}},
  \bibinfo{journal}{Phys. Rev. Lett.} \textbf{\bibinfo{volume}{92}},
  \bibinfo{pages}{173003} (\bibinfo{year}{2004}).

\bibitem[{\citenamefont{Stock et~al.}(2005)\citenamefont{Stock, Hadzibabic,
  Battelier, Cheneau, and Dalibard}}]{Stock2005}
\bibinfo{author}{\bibfnamefont{S.}~\bibnamefont{Stock}},
  \bibinfo{author}{\bibfnamefont{Z.}~\bibnamefont{Hadzibabic}},
  \bibinfo{author}{\bibfnamefont{B.}~\bibnamefont{Battelier}},
  \bibinfo{author}{\bibfnamefont{M.}~\bibnamefont{Cheneau}}, \bibnamefont{and}
  \bibinfo{author}{\bibfnamefont{J.}~\bibnamefont{Dalibard}},
  \bibinfo{journal}{Phys. Rev. Lett.} \textbf{\bibinfo{volume}{95}},
  \bibinfo{pages}{190403} (\bibinfo{year}{2005}).

\bibitem[{\citenamefont{{Smith} et~al.}(2005)\citenamefont{{Smith},
  {Heathcote}, {Hechenblaikner}, {Nugent}, and {Foot}}}]{Smith2005}
\bibinfo{author}{\bibfnamefont{N.~L.} \bibnamefont{{Smith}}},
  \bibinfo{author}{\bibfnamefont{W.~H.} \bibnamefont{{Heathcote}}},
  \bibinfo{author}{\bibfnamefont{G.}~\bibnamefont{{Hechenblaikner}}},
  \bibinfo{author}{\bibfnamefont{E.}~\bibnamefont{{Nugent}}}, \bibnamefont{and}
  \bibinfo{author}{\bibfnamefont{C.~J.} \bibnamefont{{Foot}}},
  \bibinfo{journal}{Journal of Physics B Atomic Molecular Physics}
  \textbf{\bibinfo{volume}{38}}, \bibinfo{pages}{223} (\bibinfo{year}{2005}).

\bibitem[{\citenamefont{{Hadzibabic} et~al.}(2006)\citenamefont{{Hadzibabic},
  {Kr{\"u}ger}, {Cheneau}, {Battelier}, and {Dalibard}}}]{Hadzibabic2006}
\bibinfo{author}{\bibfnamefont{Z.}~\bibnamefont{{Hadzibabic}}},
  \bibinfo{author}{\bibfnamefont{P.}~\bibnamefont{{Kr{\"u}ger}}},
  \bibinfo{author}{\bibfnamefont{M.}~\bibnamefont{{Cheneau}}},
  \bibinfo{author}{\bibfnamefont{B.}~\bibnamefont{{Battelier}}},
  \bibnamefont{and}
  \bibinfo{author}{\bibfnamefont{J.}~\bibnamefont{{Dalibard}}},
  \bibinfo{journal}{\nat} \textbf{\bibinfo{volume}{441}}, \bibinfo{pages}{1118}
  (\bibinfo{year}{2006}).

\bibitem[{\citenamefont{Schweikhard et~al.}(2007)\citenamefont{Schweikhard,
  Tung, and Cornell}}]{Schweikhard2007}
\bibinfo{author}{\bibfnamefont{V.}~\bibnamefont{Schweikhard}},
  \bibinfo{author}{\bibfnamefont{S.}~\bibnamefont{Tung}}, \bibnamefont{and}
  \bibinfo{author}{\bibfnamefont{E.~A.} \bibnamefont{Cornell}},
  \bibinfo{journal}{Phys. Rev. Lett.} \textbf{\bibinfo{volume}{99}},
  \bibinfo{pages}{030401} (\bibinfo{year}{2007}).

\bibitem[{\citenamefont{Kr\"uger et~al.}(2007)\citenamefont{Kr\"uger,
  Hadzibabic, and Dalibard}}]{Kruger2007}
\bibinfo{author}{\bibfnamefont{P.}~\bibnamefont{Kr\"uger}},
  \bibinfo{author}{\bibfnamefont{Z.}~\bibnamefont{Hadzibabic}},
  \bibnamefont{and} \bibinfo{author}{\bibfnamefont{J.}~\bibnamefont{Dalibard}},
  \bibinfo{journal}{Phys. Rev. Lett.} \textbf{\bibinfo{volume}{99}},
  \bibinfo{pages}{040402} (\bibinfo{year}{2007}).

\bibitem[{\citenamefont{{Hadzibabic} et~al.}(2008)\citenamefont{{Hadzibabic},
  {Kr{\"u}ger}, {Cheneau}, {Rath}, and {Dalibard}}}]{Rath2008}
\bibinfo{author}{\bibfnamefont{Z.}~\bibnamefont{{Hadzibabic}}},
  \bibinfo{author}{\bibfnamefont{P.}~\bibnamefont{{Kr{\"u}ger}}},
  \bibinfo{author}{\bibfnamefont{M.}~\bibnamefont{{Cheneau}}},
  \bibinfo{author}{\bibfnamefont{S.~P.} \bibnamefont{{Rath}}},
  \bibnamefont{and}
  \bibinfo{author}{\bibfnamefont{J.}~\bibnamefont{{Dalibard}}},
  \bibinfo{journal}{New Journal of Physics} \textbf{\bibinfo{volume}{10}},
  \bibinfo{pages}{045006} (\bibinfo{year}{2008}), \eprint{0712.1265}.

\bibitem[{\citenamefont{Clad\'e et~al.}(2009)\citenamefont{Clad\'e, Ryu,
  Ramanathan, Helmerson, and Phillips}}]{Clade2009}
\bibinfo{author}{\bibfnamefont{P.}~\bibnamefont{Clad\'e}},
  \bibinfo{author}{\bibfnamefont{C.}~\bibnamefont{Ryu}},
  \bibinfo{author}{\bibfnamefont{A.}~\bibnamefont{Ramanathan}},
  \bibinfo{author}{\bibfnamefont{K.}~\bibnamefont{Helmerson}},
  \bibnamefont{and} \bibinfo{author}{\bibfnamefont{W.~D.}
  \bibnamefont{Phillips}}, \bibinfo{journal}{Phys. Rev. Lett.}
  \textbf{\bibinfo{volume}{102}}, \bibinfo{pages}{170401}
  (\bibinfo{year}{2009}).

\bibitem[{\citenamefont{Rath et~al.}(2010)\citenamefont{Rath, Yefsah, G\"unter,
  Cheneau, Desbuquois, Holzmann, Krauth, and Dalibard}}]{Rath2010}
\bibinfo{author}{\bibfnamefont{S.~P.} \bibnamefont{Rath}},
  \bibinfo{author}{\bibfnamefont{T.}~\bibnamefont{Yefsah}},
  \bibinfo{author}{\bibfnamefont{K.~J.} \bibnamefont{G\"unter}},
  \bibinfo{author}{\bibfnamefont{M.}~\bibnamefont{Cheneau}},
  \bibinfo{author}{\bibfnamefont{R.}~\bibnamefont{Desbuquois}},
  \bibinfo{author}{\bibfnamefont{M.}~\bibnamefont{Holzmann}},
  \bibinfo{author}{\bibfnamefont{W.}~\bibnamefont{Krauth}}, \bibnamefont{and}
  \bibinfo{author}{\bibfnamefont{J.}~\bibnamefont{Dalibard}},
  \bibinfo{journal}{Phys. Rev. A} \textbf{\bibinfo{volume}{82}},
  \bibinfo{pages}{013609} (\bibinfo{year}{2010}).

\bibitem[{\citenamefont{Tung et~al.}(2010)\citenamefont{Tung, Lamporesi,
  Lobser, Xia, and Cornell}}]{Tung2010}
\bibinfo{author}{\bibfnamefont{S.}~\bibnamefont{Tung}},
  \bibinfo{author}{\bibfnamefont{G.}~\bibnamefont{Lamporesi}},
  \bibinfo{author}{\bibfnamefont{D.}~\bibnamefont{Lobser}},
  \bibinfo{author}{\bibfnamefont{L.}~\bibnamefont{Xia}}, \bibnamefont{and}
  \bibinfo{author}{\bibfnamefont{E.~A.} \bibnamefont{Cornell}},
  \bibinfo{journal}{Phys. Rev. Lett.} \textbf{\bibinfo{volume}{105}},
  \bibinfo{pages}{230408} (\bibinfo{year}{2010}).

\bibitem[{\citenamefont{Hung et~al.}(2011)\citenamefont{Hung, Zhang, Gemelke,
  and Chin}}]{Hung2011}
\bibinfo{author}{\bibfnamefont{C.-L.} \bibnamefont{Hung}},
  \bibinfo{author}{\bibfnamefont{X.}~\bibnamefont{Zhang}},
  \bibinfo{author}{\bibfnamefont{N.}~\bibnamefont{Gemelke}}, \bibnamefont{and}
  \bibinfo{author}{\bibfnamefont{C.}~\bibnamefont{Chin}},
  \bibinfo{journal}{Nature (adv. online)}  (\bibinfo{year}{2011}).

\bibitem[{\citenamefont{Moritz et~al.}(2003)\citenamefont{Moritz, St\"oferle,
  K\"ohl, and Esslinger}}]{Moritz2003}
\bibinfo{author}{\bibfnamefont{H.}~\bibnamefont{Moritz}},
  \bibinfo{author}{\bibfnamefont{T.}~\bibnamefont{St\"oferle}},
  \bibinfo{author}{\bibfnamefont{M.}~\bibnamefont{K\"ohl}}, \bibnamefont{and}
  \bibinfo{author}{\bibfnamefont{T.}~\bibnamefont{Esslinger}},
  \bibinfo{journal}{Phys. Rev. Lett.} \textbf{\bibinfo{volume}{91}},
  \bibinfo{pages}{250402} (\bibinfo{year}{2003}).

\bibitem[{\citenamefont{{Paredes} et~al.}(2004)\citenamefont{{Paredes},
  {Widera}, {Murg}, {Mandel}, {F{\"o}lling}, {Cirac}, {Shlyapnikov},
  {H{\"a}nsch}, and {Bloch}}}]{Paredes2004}
\bibinfo{author}{\bibfnamefont{B.}~\bibnamefont{{Paredes}}},
  \bibinfo{author}{\bibfnamefont{A.}~\bibnamefont{{Widera}}},
  \bibinfo{author}{\bibfnamefont{V.}~\bibnamefont{{Murg}}},
  \bibinfo{author}{\bibfnamefont{O.}~\bibnamefont{{Mandel}}},
  \bibinfo{author}{\bibfnamefont{S.}~\bibnamefont{{F{\"o}lling}}},
  \bibinfo{author}{\bibfnamefont{I.}~\bibnamefont{{Cirac}}},
  \bibinfo{author}{\bibfnamefont{G.~V.} \bibnamefont{{Shlyapnikov}}},
  \bibinfo{author}{\bibfnamefont{T.~W.} \bibnamefont{{H{\"a}nsch}}},
  \bibnamefont{and} \bibinfo{author}{\bibfnamefont{I.}~\bibnamefont{{Bloch}}},
  \bibinfo{journal}{\nat} \textbf{\bibinfo{volume}{429}}, \bibinfo{pages}{277}
  (\bibinfo{year}{2004}).

\bibitem[{\citenamefont{{Kinoshita} et~al.}(2004)\citenamefont{{Kinoshita},
  {Wenger}, and {Weiss}}}]{Kinoshita2004}
\bibinfo{author}{\bibfnamefont{T.}~\bibnamefont{{Kinoshita}}},
  \bibinfo{author}{\bibfnamefont{T.}~\bibnamefont{{Wenger}}}, \bibnamefont{and}
  \bibinfo{author}{\bibfnamefont{D.~S.} \bibnamefont{{Weiss}}},
  \bibinfo{journal}{Science} \textbf{\bibinfo{volume}{305}},
  \bibinfo{pages}{1125} (\bibinfo{year}{2004}).

\bibitem[{\citenamefont{Cacciapuoti et~al.}(2003)\citenamefont{Cacciapuoti,
  Hellweg, Kottke, Schulte, Ertmer, Arlt, Sengstock, Santos, and
  Lewenstein}}]{Cacciapuoti2003}
\bibinfo{author}{\bibfnamefont{L.}~\bibnamefont{Cacciapuoti}},
  \bibinfo{author}{\bibfnamefont{D.}~\bibnamefont{Hellweg}},
  \bibinfo{author}{\bibfnamefont{M.}~\bibnamefont{Kottke}},
  \bibinfo{author}{\bibfnamefont{T.}~\bibnamefont{Schulte}},
  \bibinfo{author}{\bibfnamefont{W.}~\bibnamefont{Ertmer}},
  \bibinfo{author}{\bibfnamefont{J.~J.} \bibnamefont{Arlt}},
  \bibinfo{author}{\bibfnamefont{K.}~\bibnamefont{Sengstock}},
  \bibinfo{author}{\bibfnamefont{L.}~\bibnamefont{Santos}}, \bibnamefont{and}
  \bibinfo{author}{\bibfnamefont{M.}~\bibnamefont{Lewenstein}},
  \bibinfo{journal}{Phys. Rev. A} \textbf{\bibinfo{volume}{68}},
  \bibinfo{pages}{053612} (\bibinfo{year}{2003}).

\bibitem[{\citenamefont{Trebbia et~al.}(2006)\citenamefont{Trebbia, Esteve,
  Westbrook, and Bouchoule}}]{Trebbia2006}
\bibinfo{author}{\bibfnamefont{J.-B.} \bibnamefont{Trebbia}},
  \bibinfo{author}{\bibfnamefont{J.}~\bibnamefont{Esteve}},
  \bibinfo{author}{\bibfnamefont{C.~I.} \bibnamefont{Westbrook}},
  \bibnamefont{and}
  \bibinfo{author}{\bibfnamefont{I.}~\bibnamefont{Bouchoule}},
  \bibinfo{journal}{Phys. Rev. Lett.} \textbf{\bibinfo{volume}{97}},
  \bibinfo{pages}{250403} (\bibinfo{year}{2006}).

\bibitem[{\citenamefont{van Amerongen et~al.}(2008)\citenamefont{van Amerongen,
  van Es, Wicke, Kheruntsyan, and van Druten}}]{vanAmerongen2008}
\bibinfo{author}{\bibfnamefont{A.~H.} \bibnamefont{van Amerongen}},
  \bibinfo{author}{\bibfnamefont{J.~J.~P.} \bibnamefont{van Es}},
  \bibinfo{author}{\bibfnamefont{P.}~\bibnamefont{Wicke}},
  \bibinfo{author}{\bibfnamefont{K.~V.} \bibnamefont{Kheruntsyan}},
  \bibnamefont{and} \bibinfo{author}{\bibfnamefont{N.~J.} \bibnamefont{van
  Druten}}, \bibinfo{journal}{Phys. Rev. Lett.} \textbf{\bibinfo{volume}{100}},
  \bibinfo{pages}{090402} (\bibinfo{year}{2008}).

\bibitem[{\citenamefont{Armijo et~al.}(2010)\citenamefont{Armijo, Jacqmin,
  Kheruntsyan, and Bouchoule}}]{Armijo2010}
\bibinfo{author}{\bibfnamefont{J.}~\bibnamefont{Armijo}},
  \bibinfo{author}{\bibfnamefont{T.}~\bibnamefont{Jacqmin}},
  \bibinfo{author}{\bibfnamefont{K.~V.} \bibnamefont{Kheruntsyan}},
  \bibnamefont{and}
  \bibinfo{author}{\bibfnamefont{I.}~\bibnamefont{Bouchoule}},
  \bibinfo{journal}{Phys. Rev. Lett.} \textbf{\bibinfo{volume}{105}},
  \bibinfo{pages}{230402} (\bibinfo{year}{2010}).

\bibitem[{\citenamefont{Armijo et~al.}(2011)\citenamefont{Armijo, Jacqmin,
  Kheruntsyan, and Bouchoule}}]{Armijo2011}
\bibinfo{author}{\bibfnamefont{J.}~\bibnamefont{Armijo}},
  \bibinfo{author}{\bibfnamefont{T.}~\bibnamefont{Jacqmin}},
  \bibinfo{author}{\bibfnamefont{K.}~\bibnamefont{Kheruntsyan}},
  \bibnamefont{and}
  \bibinfo{author}{\bibfnamefont{I.}~\bibnamefont{Bouchoule}},
  \bibinfo{journal}{Phys. Rev. A} \textbf{\bibinfo{volume}{83}},
  \bibinfo{pages}{021605} (\bibinfo{year}{2011}).

\bibitem[{\citenamefont{Est\`eve et~al.}(2006)\citenamefont{Est\`eve, Trebbia,
  Schumm, Aspect, Westbrook, and Bouchoule}}]{Esteve2006}
\bibinfo{author}{\bibfnamefont{J.}~\bibnamefont{Est\`eve}},
  \bibinfo{author}{\bibfnamefont{J.-B.} \bibnamefont{Trebbia}},
  \bibinfo{author}{\bibfnamefont{T.}~\bibnamefont{Schumm}},
  \bibinfo{author}{\bibfnamefont{A.}~\bibnamefont{Aspect}},
  \bibinfo{author}{\bibfnamefont{C.~I.} \bibnamefont{Westbrook}},
  \bibnamefont{and}
  \bibinfo{author}{\bibfnamefont{I.}~\bibnamefont{Bouchoule}},
  \bibinfo{journal}{Phys. Rev. Lett.} \textbf{\bibinfo{volume}{96}},
  \bibinfo{pages}{130403} (\bibinfo{year}{2006}).

\bibitem[{\citenamefont{Dettmer et~al.}(2001)\citenamefont{Dettmer, Hellweg,
  Ryytty, Arlt, Ertmer, Sengstock, Petrov, Shlyapnikov, Kreutzmann, Santos
  et~al.}}]{Dettmer2001}
\bibinfo{author}{\bibfnamefont{S.}~\bibnamefont{Dettmer}},
  \bibinfo{author}{\bibfnamefont{D.}~\bibnamefont{Hellweg}},
  \bibinfo{author}{\bibfnamefont{P.}~\bibnamefont{Ryytty}},
  \bibinfo{author}{\bibfnamefont{J.~J.} \bibnamefont{Arlt}},
  \bibinfo{author}{\bibfnamefont{W.}~\bibnamefont{Ertmer}},
  \bibinfo{author}{\bibfnamefont{K.}~\bibnamefont{Sengstock}},
  \bibinfo{author}{\bibfnamefont{D.~S.} \bibnamefont{Petrov}},
  \bibinfo{author}{\bibfnamefont{G.~V.} \bibnamefont{Shlyapnikov}},
  \bibinfo{author}{\bibfnamefont{H.}~\bibnamefont{Kreutzmann}},
  \bibinfo{author}{\bibfnamefont{L.}~\bibnamefont{Santos}},
  \bibnamefont{et~al.}, \bibinfo{journal}{Phys. Rev. Lett.}
  \textbf{\bibinfo{volume}{87}}, \bibinfo{pages}{160406}
  (\bibinfo{year}{2001}).

\bibitem[{\citenamefont{Hellweg et~al.}(2003)\citenamefont{Hellweg,
  Cacciapuoti, Kottke, Schulte, Sengstock, Ertmer, and Arlt}}]{Hellweg2003}
\bibinfo{author}{\bibfnamefont{D.}~\bibnamefont{Hellweg}},
  \bibinfo{author}{\bibfnamefont{L.}~\bibnamefont{Cacciapuoti}},
  \bibinfo{author}{\bibfnamefont{M.}~\bibnamefont{Kottke}},
  \bibinfo{author}{\bibfnamefont{T.}~\bibnamefont{Schulte}},
  \bibinfo{author}{\bibfnamefont{K.}~\bibnamefont{Sengstock}},
  \bibinfo{author}{\bibfnamefont{W.}~\bibnamefont{Ertmer}}, \bibnamefont{and}
  \bibinfo{author}{\bibfnamefont{J.~J.} \bibnamefont{Arlt}},
  \bibinfo{journal}{Phys. Rev. Lett.} \textbf{\bibinfo{volume}{91}},
  \bibinfo{pages}{010406} (\bibinfo{year}{2003}).

\bibitem[{\citenamefont{Gerbier et~al.}(2003)\citenamefont{Gerbier, Thywissen,
  Richard, Hugbart, Bouyer, and Aspect}}]{Gerbier2003}
\bibinfo{author}{\bibfnamefont{F.}~\bibnamefont{Gerbier}},
  \bibinfo{author}{\bibfnamefont{J.~H.} \bibnamefont{Thywissen}},
  \bibinfo{author}{\bibfnamefont{S.}~\bibnamefont{Richard}},
  \bibinfo{author}{\bibfnamefont{M.}~\bibnamefont{Hugbart}},
  \bibinfo{author}{\bibfnamefont{P.}~\bibnamefont{Bouyer}}, \bibnamefont{and}
  \bibinfo{author}{\bibfnamefont{A.}~\bibnamefont{Aspect}},
  \bibinfo{journal}{Phys. Rev. A} \textbf{\bibinfo{volume}{67}},
  \bibinfo{pages}{051602} (\bibinfo{year}{2003}).

\bibitem[{\citenamefont{Richard et~al.}(2003)\citenamefont{Richard, Gerbier,
  Thywissen, Hugbart, Bouyer, and Aspect}}]{Richard2003}
\bibinfo{author}{\bibfnamefont{S.}~\bibnamefont{Richard}},
  \bibinfo{author}{\bibfnamefont{F.}~\bibnamefont{Gerbier}},
  \bibinfo{author}{\bibfnamefont{J.~H.} \bibnamefont{Thywissen}},
  \bibinfo{author}{\bibfnamefont{M.}~\bibnamefont{Hugbart}},
  \bibinfo{author}{\bibfnamefont{P.}~\bibnamefont{Bouyer}}, \bibnamefont{and}
  \bibinfo{author}{\bibfnamefont{A.}~\bibnamefont{Aspect}},
  \bibinfo{journal}{Phys. Rev. Lett.} \textbf{\bibinfo{volume}{91}},
  \bibinfo{pages}{010405} (\bibinfo{year}{2003}).

\bibitem[{\citenamefont{Manz et~al.}(2010)\citenamefont{Manz, B\"ucker, Betz,
  Koller, Hofferberth, Mazets, Imambekov, Demler, Perrin, Schmiedmayer
  et~al.}}]{Manz2010}
\bibinfo{author}{\bibfnamefont{S.}~\bibnamefont{Manz}},
  \bibinfo{author}{\bibfnamefont{R.}~\bibnamefont{B\"ucker}},
  \bibinfo{author}{\bibfnamefont{T.}~\bibnamefont{Betz}},
  \bibinfo{author}{\bibfnamefont{C.}~\bibnamefont{Koller}},
  \bibinfo{author}{\bibfnamefont{S.}~\bibnamefont{Hofferberth}},
  \bibinfo{author}{\bibfnamefont{I.~E.} \bibnamefont{Mazets}},
  \bibinfo{author}{\bibfnamefont{A.}~\bibnamefont{Imambekov}},
  \bibinfo{author}{\bibfnamefont{E.}~\bibnamefont{Demler}},
  \bibinfo{author}{\bibfnamefont{A.}~\bibnamefont{Perrin}},
  \bibinfo{author}{\bibfnamefont{J.}~\bibnamefont{Schmiedmayer}},
  \bibnamefont{et~al.}, \bibinfo{journal}{Phys. Rev. A}
  \textbf{\bibinfo{volume}{81}}, \bibinfo{pages}{031610}
  (\bibinfo{year}{2010}).

\bibitem[{\citenamefont{Gustavson et~al.}(1997)\citenamefont{Gustavson, Bouyer,
  and Kasevich}}]{Gustavson1997}
\bibinfo{author}{\bibfnamefont{T.~L.} \bibnamefont{Gustavson}},
  \bibinfo{author}{\bibfnamefont{P.}~\bibnamefont{Bouyer}}, \bibnamefont{and}
  \bibinfo{author}{\bibfnamefont{M.~A.} \bibnamefont{Kasevich}},
  \bibinfo{journal}{Phys. Rev. Lett.} \textbf{\bibinfo{volume}{78}},
  \bibinfo{pages}{2046} (\bibinfo{year}{1997}).

\bibitem[{\citenamefont{Hinds et~al.}(2001)\citenamefont{Hinds, Vale, and
  Boshier}}]{Hinds2001}
\bibinfo{author}{\bibfnamefont{E.~A.} \bibnamefont{Hinds}},
  \bibinfo{author}{\bibfnamefont{C.~J.} \bibnamefont{Vale}}, \bibnamefont{and}
  \bibinfo{author}{\bibfnamefont{M.~G.} \bibnamefont{Boshier}},
  \bibinfo{journal}{Phys. Rev. Lett.} \textbf{\bibinfo{volume}{86}},
  \bibinfo{pages}{1462} (\bibinfo{year}{2001}).

\bibitem[{\citenamefont{{Schumm} et~al.}(2005)\citenamefont{{Schumm},
  {Hofferberth}, {Andersson}, {Wildermuth}, {Groth}, {Bar-Joseph},
  {Schmiedmayer}, and {Kr{\"u}ger}}}]{Schumm2005}
\bibinfo{author}{\bibfnamefont{T.}~\bibnamefont{{Schumm}}},
  \bibinfo{author}{\bibfnamefont{S.}~\bibnamefont{{Hofferberth}}},
  \bibinfo{author}{\bibfnamefont{L.~M.} \bibnamefont{{Andersson}}},
  \bibinfo{author}{\bibfnamefont{S.}~\bibnamefont{{Wildermuth}}},
  \bibinfo{author}{\bibfnamefont{S.}~\bibnamefont{{Groth}}},
  \bibinfo{author}{\bibfnamefont{I.}~\bibnamefont{{Bar-Joseph}}},
  \bibinfo{author}{\bibfnamefont{J.}~\bibnamefont{{Schmiedmayer}}},
  \bibnamefont{and}
  \bibinfo{author}{\bibfnamefont{P.}~\bibnamefont{{Kr{\"u}ger}}},
  \bibinfo{journal}{Nature Physics} \textbf{\bibinfo{volume}{1}},
  \bibinfo{pages}{57} (\bibinfo{year}{2005}).

\bibitem[{\citenamefont{{Hofferberth} et~al.}(2006)\citenamefont{{Hofferberth},
  {Lesanovsky}, {Fischer}, {Verdu}, and {Schmiedmayer}}}]{Hofferberth2006}
\bibinfo{author}{\bibfnamefont{S.}~\bibnamefont{{Hofferberth}}},
  \bibinfo{author}{\bibfnamefont{I.}~\bibnamefont{{Lesanovsky}}},
  \bibinfo{author}{\bibfnamefont{B.}~\bibnamefont{{Fischer}}},
  \bibinfo{author}{\bibfnamefont{J.}~\bibnamefont{{Verdu}}}, \bibnamefont{and}
  \bibinfo{author}{\bibfnamefont{J.}~\bibnamefont{{Schmiedmayer}}},
  \bibinfo{journal}{Nature Physics} \textbf{\bibinfo{volume}{2}},
  \bibinfo{pages}{710} (\bibinfo{year}{2006}).

\bibitem[{\citenamefont{{Fixler} et~al.}(2007)\citenamefont{{Fixler}, {Foster},
  {McGuirk}, and {Kasevich}}}]{Fixler2007}
\bibinfo{author}{\bibfnamefont{J.~B.} \bibnamefont{{Fixler}}},
  \bibinfo{author}{\bibfnamefont{G.~T.} \bibnamefont{{Foster}}},
  \bibinfo{author}{\bibfnamefont{J.~M.} \bibnamefont{{McGuirk}}},
  \bibnamefont{and} \bibinfo{author}{\bibfnamefont{M.~A.}
  \bibnamefont{{Kasevich}}}, \bibinfo{journal}{Science}
  \textbf{\bibinfo{volume}{315}}, \bibinfo{pages}{74} (\bibinfo{year}{2007}).

\bibitem[{\citenamefont{Jo et~al.}(2007)\citenamefont{Jo, Shin, Will, Pasquini,
  Saba, Ketterle, Pritchard, Vengalattore, and Prentiss}}]{Jo2007b}
\bibinfo{author}{\bibfnamefont{G.-B.} \bibnamefont{Jo}},
  \bibinfo{author}{\bibfnamefont{Y.}~\bibnamefont{Shin}},
  \bibinfo{author}{\bibfnamefont{S.}~\bibnamefont{Will}},
  \bibinfo{author}{\bibfnamefont{T.~A.} \bibnamefont{Pasquini}},
  \bibinfo{author}{\bibfnamefont{M.}~\bibnamefont{Saba}},
  \bibinfo{author}{\bibfnamefont{W.}~\bibnamefont{Ketterle}},
  \bibinfo{author}{\bibfnamefont{D.~E.} \bibnamefont{Pritchard}},
  \bibinfo{author}{\bibfnamefont{M.}~\bibnamefont{Vengalattore}},
  \bibnamefont{and} \bibinfo{author}{\bibfnamefont{M.}~\bibnamefont{Prentiss}},
  \bibinfo{journal}{Phys. Rev. Lett.} \textbf{\bibinfo{volume}{98}},
  \bibinfo{pages}{030407} (\bibinfo{year}{2007}).

\bibitem[{\citenamefont{{Hofferberth} et~al.}(2007)\citenamefont{{Hofferberth},
  {Lesanovsky}, {Fischer}, {Schumm}, and {Schmiedmayer}}}]{Hofferberth2007}
\bibinfo{author}{\bibfnamefont{S.}~\bibnamefont{{Hofferberth}}},
  \bibinfo{author}{\bibfnamefont{I.}~\bibnamefont{{Lesanovsky}}},
  \bibinfo{author}{\bibfnamefont{B.}~\bibnamefont{{Fischer}}},
  \bibinfo{author}{\bibfnamefont{T.}~\bibnamefont{{Schumm}}}, \bibnamefont{and}
  \bibinfo{author}{\bibfnamefont{J.}~\bibnamefont{{Schmiedmayer}}},
  \bibinfo{journal}{\nat} \textbf{\bibinfo{volume}{449}}, \bibinfo{pages}{324}
  (\bibinfo{year}{2007}).

\bibitem[{\citenamefont{{Gross} et~al.}(2010)\citenamefont{{Gross}, {Zibold},
  {Nicklas}, {Est{\`e}ve}, and {Oberthaler}}}]{Gross2010}
\bibinfo{author}{\bibfnamefont{C.}~\bibnamefont{{Gross}}},
  \bibinfo{author}{\bibfnamefont{T.}~\bibnamefont{{Zibold}}},
  \bibinfo{author}{\bibfnamefont{E.}~\bibnamefont{{Nicklas}}},
  \bibinfo{author}{\bibfnamefont{J.}~\bibnamefont{{Est{\`e}ve}}},
  \bibnamefont{and} \bibinfo{author}{\bibfnamefont{M.~K.}
  \bibnamefont{{Oberthaler}}}, \bibinfo{journal}{\nat}
  \textbf{\bibinfo{volume}{464}}, \bibinfo{pages}{1165} (\bibinfo{year}{2010}),
  \eprint{1009.2374}.

\bibitem[{\citenamefont{Baumg\"artner et~al.}(2010)\citenamefont{Baumg\"artner,
  Sewell, Eriksson, Llorente-Garcia, Dingjan, Cotter, and
  Hinds}}]{Baumgartner2010}
\bibinfo{author}{\bibfnamefont{F.}~\bibnamefont{Baumg\"artner}},
  \bibinfo{author}{\bibfnamefont{R.~J.} \bibnamefont{Sewell}},
  \bibinfo{author}{\bibfnamefont{S.}~\bibnamefont{Eriksson}},
  \bibinfo{author}{\bibfnamefont{I.}~\bibnamefont{Llorente-Garcia}},
  \bibinfo{author}{\bibfnamefont{J.}~\bibnamefont{Dingjan}},
  \bibinfo{author}{\bibfnamefont{J.~P.} \bibnamefont{Cotter}},
  \bibnamefont{and} \bibinfo{author}{\bibfnamefont{E.~A.} \bibnamefont{Hinds}},
  \bibinfo{journal}{Phys. Rev. Lett.} \textbf{\bibinfo{volume}{105}},
  \bibinfo{pages}{243003} (\bibinfo{year}{2010}).

\bibitem[{\citenamefont{{Hinds} and {Hughes}}(1999)}]{Hinds1999}
\bibinfo{author}{\bibfnamefont{E.~A.} \bibnamefont{{Hinds}}} \bibnamefont{and}
  \bibinfo{author}{\bibfnamefont{I.~G.} \bibnamefont{{Hughes}}},
  \bibinfo{journal}{Journal of Physics D Applied Physics}
  \textbf{\bibinfo{volume}{32}}, \bibinfo{pages}{119} (\bibinfo{year}{1999}).

\bibitem[{\citenamefont{Ott et~al.}(2001)\citenamefont{Ott, Fortagh,
  Schlotterbeck, Grossmann, and Zimmermann}}]{Ott2001}
\bibinfo{author}{\bibfnamefont{H.}~\bibnamefont{Ott}},
  \bibinfo{author}{\bibfnamefont{J.}~\bibnamefont{Fortagh}},
  \bibinfo{author}{\bibfnamefont{G.}~\bibnamefont{Schlotterbeck}},
  \bibinfo{author}{\bibfnamefont{A.}~\bibnamefont{Grossmann}},
  \bibnamefont{and}
  \bibinfo{author}{\bibfnamefont{C.}~\bibnamefont{Zimmermann}},
  \bibinfo{journal}{Phys. Rev. Lett.} \textbf{\bibinfo{volume}{87}},
  \bibinfo{pages}{230401} (\bibinfo{year}{2001}).

\bibitem[{\citenamefont{{H{\"a}nsel} et~al.}(2001)\citenamefont{{H{\"a}nsel},
  {Hommelhoff}, {H{\"a}nsch}, and {Reichel}}}]{Hansel2001}
\bibinfo{author}{\bibfnamefont{W.}~\bibnamefont{{H{\"a}nsel}}},
  \bibinfo{author}{\bibfnamefont{P.}~\bibnamefont{{Hommelhoff}}},
  \bibinfo{author}{\bibfnamefont{T.~W.} \bibnamefont{{H{\"a}nsch}}},
  \bibnamefont{and}
  \bibinfo{author}{\bibfnamefont{J.}~\bibnamefont{{Reichel}}},
  \bibinfo{journal}{\nat} \textbf{\bibinfo{volume}{413}}, \bibinfo{pages}{498}
  (\bibinfo{year}{2001}).

\bibitem[{\citenamefont{{Folman} et~al.}(2002)\citenamefont{{Folman},
  {Kr{\"u}ger}, {Schmiedmayer}, {Denschlag}, and {Henkel}}}]{Folman2002}
\bibinfo{author}{\bibfnamefont{R.}~\bibnamefont{{Folman}}},
  \bibinfo{author}{\bibfnamefont{P.}~\bibnamefont{{Kr{\"u}ger}}},
  \bibinfo{author}{\bibfnamefont{J.}~\bibnamefont{{Schmiedmayer}}},
  \bibinfo{author}{\bibfnamefont{J.}~\bibnamefont{{Denschlag}}},
  \bibnamefont{and} \bibinfo{author}{\bibfnamefont{C.}~\bibnamefont{{Henkel}}},
  \bibinfo{journal}{Advances in Atomic and Molecular Physics}
  \textbf{\bibinfo{volume}{48}}, \bibinfo{pages}{263} (\bibinfo{year}{2002}),
  \eprint{0805.2613}.

\bibitem[{\citenamefont{Fort\'agh and Zimmermann}(2007)}]{Fortagh2007}
\bibinfo{author}{\bibfnamefont{J.}~\bibnamefont{Fort\'agh}} \bibnamefont{and}
  \bibinfo{author}{\bibfnamefont{C.}~\bibnamefont{Zimmermann}},
  \bibinfo{journal}{Rev. Mod. Phys.} \textbf{\bibinfo{volume}{79}},
  \bibinfo{pages}{235} (\bibinfo{year}{2007}).

\bibitem[{\citenamefont{Reichel and Vuletic}(2011)}]{Atomchips2011}
\bibinfo{author}{\bibfnamefont{J.}~\bibnamefont{Reichel}} \bibnamefont{and}
  \bibinfo{author}{\bibfnamefont{V.~E.} \bibnamefont{Vuletic}},
  \emph{\bibinfo{title}{Atom Chips}} (\bibinfo{publisher}{WILEY-VCH Verlag GmbH
  \& Co. KGaA}, \bibinfo{address}{Weinheim}, \bibinfo{year}{2011}).

\bibitem[{\citenamefont{Petrov et~al.}(2000)\citenamefont{Petrov, Shlyapnikov,
  and Walraven}}]{Petrov2000}
\bibinfo{author}{\bibfnamefont{D.~S.} \bibnamefont{Petrov}},
  \bibinfo{author}{\bibfnamefont{G.~V.} \bibnamefont{Shlyapnikov}},
  \bibnamefont{and} \bibinfo{author}{\bibfnamefont{J.~T.~M.}
  \bibnamefont{Walraven}}, \bibinfo{journal}{Phys. Rev. Lett.}
  \textbf{\bibinfo{volume}{85}}, \bibinfo{pages}{3745} (\bibinfo{year}{2000}).

\bibitem[{\citenamefont{{Girardeau}}(1960)}]{Girardeau1960}
\bibinfo{author}{\bibfnamefont{M.}~\bibnamefont{{Girardeau}}},
  \bibinfo{journal}{Journal of Mathematical Physics}
  \textbf{\bibinfo{volume}{1}}, \bibinfo{pages}{516} (\bibinfo{year}{1960}).

\bibitem[{\citenamefont{Al~Khawaja et~al.}(2003)\citenamefont{Al~Khawaja,
  Proukakis, Andersen, Romans, and Stoof}}]{AlKhawaja2003}
\bibinfo{author}{\bibfnamefont{U.}~\bibnamefont{Al~Khawaja}},
  \bibinfo{author}{\bibfnamefont{N.~P.} \bibnamefont{Proukakis}},
  \bibinfo{author}{\bibfnamefont{J.~O.} \bibnamefont{Andersen}},
  \bibinfo{author}{\bibfnamefont{M.~W.~J.} \bibnamefont{Romans}},
  \bibnamefont{and} \bibinfo{author}{\bibfnamefont{H.~T.~C.}
  \bibnamefont{Stoof}}, \bibinfo{journal}{Phys. Rev. A}
  \textbf{\bibinfo{volume}{68}}, \bibinfo{pages}{043603}
  (\bibinfo{year}{2003}).

\bibitem[{\citenamefont{Popov}(1983)}]{PopovBook}
\bibinfo{author}{\bibfnamefont{V.~N.} \bibnamefont{Popov}},
  \emph{\bibinfo{title}{Functional Integrals in Quantum Field Theory and
  Statistical Physics}} (\bibinfo{publisher}{Reidel, Dordrecht},
  \bibinfo{year}{1983}), chap.~\bibinfo{chapter}{6}.

\bibitem[{\citenamefont{Stoof}(1999)}]{Stoof1999}
\bibinfo{author}{\bibfnamefont{H.~T.~C.} \bibnamefont{Stoof}},
  \bibinfo{journal}{J. Low Temp. Phys.} \textbf{\bibinfo{volume}{114}},
  \bibinfo{pages}{11} (\bibinfo{year}{1999}).

\bibitem[{\citenamefont{Stoof and Bijlsma}(2001)}]{Stoof2001}
\bibinfo{author}{\bibfnamefont{H.~T.~C.} \bibnamefont{Stoof}} \bibnamefont{and}
  \bibinfo{author}{\bibfnamefont{M.~J.} \bibnamefont{Bijlsma}},
  \bibinfo{journal}{J. Low Temp. Phys.} \textbf{\bibinfo{volume}{124}},
  \bibinfo{pages}{431} (\bibinfo{year}{2001}).

\bibitem[{\citenamefont{Gardiner and Davis}(2003)}]{Gardiner2003}
\bibinfo{author}{\bibfnamefont{C.~W.} \bibnamefont{Gardiner}} \bibnamefont{and}
  \bibinfo{author}{\bibfnamefont{M.~J.} \bibnamefont{Davis}},
  \bibinfo{journal}{J. Phys. B} \textbf{\bibinfo{volume}{36}},
  \bibinfo{pages}{4731} (\bibinfo{year}{2003}).

\bibitem[{\citenamefont{Stringari}(1998)}]{Stringari1998}
\bibinfo{author}{\bibfnamefont{S.}~\bibnamefont{Stringari}},
  \bibinfo{journal}{Phys. Rev. A} \textbf{\bibinfo{volume}{58}},
  \bibinfo{pages}{2385} (\bibinfo{year}{1998}).

\bibitem[{\citenamefont{Petrov et~al.}(2001)\citenamefont{Petrov, Shlyapnikov,
  and Walraven}}]{Petrov2001}
\bibinfo{author}{\bibfnamefont{D.~S.} \bibnamefont{Petrov}},
  \bibinfo{author}{\bibfnamefont{G.~V.} \bibnamefont{Shlyapnikov}},
  \bibnamefont{and} \bibinfo{author}{\bibfnamefont{J.~T.~M.}
  \bibnamefont{Walraven}}, \bibinfo{journal}{Phys. Rev. Lett.}
  \textbf{\bibinfo{volume}{87}}, \bibinfo{pages}{050404}
  (\bibinfo{year}{2001}).

\bibitem[{\citenamefont{Svistunov}(1991)}]{Svistunov1991}
\bibinfo{author}{\bibfnamefont{B.}~\bibnamefont{Svistunov}},
  \bibinfo{journal}{J. Moscow Phys. Soc.} \textbf{\bibinfo{volume}{1}},
  \bibinfo{pages}{373} (\bibinfo{year}{1991}).

\bibitem[{\citenamefont{Duine and Stoof}(2001)}]{Duine2001}
\bibinfo{author}{\bibfnamefont{R.~A.} \bibnamefont{Duine}} \bibnamefont{and}
  \bibinfo{author}{\bibfnamefont{H.~T.~C.} \bibnamefont{Stoof}},
  \bibinfo{journal}{Phys. Rev. A} \textbf{\bibinfo{volume}{65}},
  \bibinfo{pages}{013603} (\bibinfo{year}{2001}).

\bibitem[{\citenamefont{Davis et~al.}(2001)\citenamefont{Davis, Morgan, and
  Burnett}}]{Davis2001}
\bibinfo{author}{\bibfnamefont{M.~J.} \bibnamefont{Davis}},
  \bibinfo{author}{\bibfnamefont{S.~A.} \bibnamefont{Morgan}},
  \bibnamefont{and} \bibinfo{author}{\bibfnamefont{K.}~\bibnamefont{Burnett}},
  \bibinfo{journal}{Phys. Rev. Lett.} \textbf{\bibinfo{volume}{87}},
  \bibinfo{pages}{160402} (\bibinfo{year}{2001}).

\bibitem[{\citenamefont{Sinatra et~al.}(2001)\citenamefont{Sinatra, Lobo, and
  Castin}}]{Sinatra2001}
\bibinfo{author}{\bibfnamefont{A.}~\bibnamefont{Sinatra}},
  \bibinfo{author}{\bibfnamefont{C.}~\bibnamefont{Lobo}}, \bibnamefont{and}
  \bibinfo{author}{\bibfnamefont{Y.}~\bibnamefont{Castin}},
  \bibinfo{journal}{Phys. Rev. Lett.} \textbf{\bibinfo{volume}{87}},
  \bibinfo{pages}{210404} (\bibinfo{year}{2001}).

\bibitem[{\citenamefont{{Goral} et~al.}(2001)\citenamefont{{Goral}, {Gajda},
  and {Rzazewski}}}]{Goral2001}
\bibinfo{author}{\bibfnamefont{K.}~\bibnamefont{{Goral}}},
  \bibinfo{author}{\bibfnamefont{M.}~\bibnamefont{{Gajda}}}, \bibnamefont{and}
  \bibinfo{author}{\bibfnamefont{K.~M.} \bibnamefont{{Rzazewski}}},
  \bibinfo{journal}{Optics Express} \textbf{\bibinfo{volume}{8}},
  \bibinfo{pages}{92} (\bibinfo{year}{2001}).

\bibitem[{\citenamefont{Penckwitt et~al.}(2002)\citenamefont{Penckwitt,
  Ballagh, and Gardiner}}]{Penckwitt2002}
\bibinfo{author}{\bibfnamefont{A.~A.} \bibnamefont{Penckwitt}},
  \bibinfo{author}{\bibfnamefont{R.~J.} \bibnamefont{Ballagh}},
  \bibnamefont{and} \bibinfo{author}{\bibfnamefont{C.~W.}
  \bibnamefont{Gardiner}}, \bibinfo{journal}{Phys. Rev. Lett.}
  \textbf{\bibinfo{volume}{89}}, \bibinfo{pages}{260402}
  (\bibinfo{year}{2002}).

\bibitem[{\citenamefont{Duine et~al.}(2004)\citenamefont{Duine, Leurs, and
  Stoof}}]{Duine2004}
\bibinfo{author}{\bibfnamefont{R.~A.} \bibnamefont{Duine}},
  \bibinfo{author}{\bibfnamefont{B.~W.~A.} \bibnamefont{Leurs}},
  \bibnamefont{and} \bibinfo{author}{\bibfnamefont{H.~T.~C.}
  \bibnamefont{Stoof}}, \bibinfo{journal}{Phys. Rev. A}
  \textbf{\bibinfo{volume}{69}}, \bibinfo{pages}{053623}
  (\bibinfo{year}{2004}).

\bibitem[{\citenamefont{Kr\"uger et~al.}(2010)\citenamefont{Kr\"uger,
  Hofferberth, Mazets, Lesanovsky, and Schmiedmayer}}]{Kruger2010}
\bibinfo{author}{\bibfnamefont{P.}~\bibnamefont{Kr\"uger}},
  \bibinfo{author}{\bibfnamefont{S.}~\bibnamefont{Hofferberth}},
  \bibinfo{author}{\bibfnamefont{I.~E.} \bibnamefont{Mazets}},
  \bibinfo{author}{\bibfnamefont{I.}~\bibnamefont{Lesanovsky}},
  \bibnamefont{and}
  \bibinfo{author}{\bibfnamefont{J.}~\bibnamefont{Schmiedmayer}},
  \bibinfo{journal}{Phys. Rev. Lett.} \textbf{\bibinfo{volume}{105}},
  \bibinfo{pages}{265302} (\bibinfo{year}{2010}).

\bibitem[{\citenamefont{Fuchs et~al.}(2003)\citenamefont{Fuchs, Leyronas, and
  Combescot}}]{Fuchs2003}
\bibinfo{author}{\bibfnamefont{J.~N.} \bibnamefont{Fuchs}},
  \bibinfo{author}{\bibfnamefont{X.}~\bibnamefont{Leyronas}}, \bibnamefont{and}
  \bibinfo{author}{\bibfnamefont{R.}~\bibnamefont{Combescot}},
  \bibinfo{journal}{Phys. Rev. A} \textbf{\bibinfo{volume}{68}},
  \bibinfo{pages}{043610} (\bibinfo{year}{2003}).

\bibitem[{\citenamefont{Gerbier}(2004)}]{Gerbier2004}
\bibinfo{author}{\bibfnamefont{F.}~\bibnamefont{Gerbier}},
  \bibinfo{journal}{Europhys. Lett.} \textbf{\bibinfo{volume}{66}},
  \bibinfo{pages}{771} (\bibinfo{year}{2004}).

\bibitem[{\citenamefont{Mu\~noz Mateo and Delgado}(2007)}]{Mateo2007}
\bibinfo{author}{\bibfnamefont{A.}~\bibnamefont{Mu\~noz Mateo}}
  \bibnamefont{and} \bibinfo{author}{\bibfnamefont{V.}~\bibnamefont{Delgado}},
  \bibinfo{journal}{Phys. Rev. A} \textbf{\bibinfo{volume}{75}},
  \bibinfo{pages}{063610} (\bibinfo{year}{2007}).

\bibitem[{\citenamefont{{Frantzeskakis}}(2010)}]{Frantzeskakis2010}
\bibinfo{author}{\bibfnamefont{D.~J.} \bibnamefont{{Frantzeskakis}}},
  \bibinfo{journal}{Journal of Physics A Mathematical General}
  \textbf{\bibinfo{volume}{43}}, \bibinfo{pages}{213001}
  (\bibinfo{year}{2010}).

\bibitem[{\citenamefont{Menotti and Stringari}(2002)}]{Menotti2002}
\bibinfo{author}{\bibfnamefont{C.}~\bibnamefont{Menotti}} \bibnamefont{and}
  \bibinfo{author}{\bibfnamefont{S.}~\bibnamefont{Stringari}},
  \bibinfo{journal}{Phys. Rev. A} \textbf{\bibinfo{volume}{66}},
  \bibinfo{pages}{043610} (\bibinfo{year}{2002}).

\bibitem[{\citenamefont{Proukakis and Jackson}(2008)}]{Proukakis2008}
\bibinfo{author}{\bibfnamefont{N.~P.} \bibnamefont{Proukakis}}
  \bibnamefont{and} \bibinfo{author}{\bibfnamefont{B.}~\bibnamefont{Jackson}},
  \bibinfo{journal}{J. Phys. B} \textbf{\bibinfo{volume}{41}},
  \bibinfo{pages}{203002} (\bibinfo{year}{2008}).

\bibitem[{\citenamefont{{Cockburn} et~al.}(2010)\citenamefont{{Cockburn},
  {Negretti}, {Proukakis}, and {Henkel}}}]{Cockburn2011}
\bibinfo{author}{\bibfnamefont{S.~P.} \bibnamefont{{Cockburn}}},
  \bibinfo{author}{\bibfnamefont{A.}~\bibnamefont{{Negretti}}},
  \bibinfo{author}{\bibfnamefont{N.~P.} \bibnamefont{{Proukakis}}},
  \bibnamefont{and} \bibinfo{author}{\bibfnamefont{C.}~\bibnamefont{{Henkel}}},
  \bibinfo{journal}{(accepted in Phys. Rev. A)}  (\bibinfo{year}{2010}),
  \eprint{arXiv:1012.1512}.

\bibitem[{\citenamefont{{Wright} et~al.}(2010)\citenamefont{{Wright},
  {Proukakis}, and {Davis}}}]{Wright2011}
\bibinfo{author}{\bibfnamefont{T.~M.} \bibnamefont{{Wright}}},
  \bibinfo{author}{\bibfnamefont{N.~P.} \bibnamefont{{Proukakis}}},
  \bibnamefont{and} \bibinfo{author}{\bibfnamefont{M.~J.}
  \bibnamefont{{Davis}}} (\bibinfo{year}{2010}), \eprint{arXiv:1011.6289}.

\bibitem[{\citenamefont{Naraschewski and Glauber}(1999)}]{Naraschewski1999}
\bibinfo{author}{\bibfnamefont{M.}~\bibnamefont{Naraschewski}}
  \bibnamefont{and} \bibinfo{author}{\bibfnamefont{R.~J.}
  \bibnamefont{Glauber}}, \bibinfo{journal}{Phys. Rev. A}
  \textbf{\bibinfo{volume}{59}}, \bibinfo{pages}{4595} (\bibinfo{year}{1999}).

\bibitem[{\citenamefont{Prokof'ev and Svistunov}(2002)}]{Prokofev2002}
\bibinfo{author}{\bibfnamefont{N.}~\bibnamefont{Prokof'ev}} \bibnamefont{and}
  \bibinfo{author}{\bibfnamefont{B.}~\bibnamefont{Svistunov}},
  \bibinfo{journal}{Phys. Rev. A} \textbf{\bibinfo{volume}{66}},
  \bibinfo{pages}{043608} (\bibinfo{year}{2002}).

\bibitem[{\citenamefont{Proukakis}(2006)}]{Proukakis2006b}
\bibinfo{author}{\bibfnamefont{N.~P.} \bibnamefont{Proukakis}},
  \bibinfo{journal}{Phys. Rev. A} \textbf{\bibinfo{volume}{74}},
  \bibinfo{pages}{053617} (\bibinfo{year}{2006}).

\bibitem[{\citenamefont{Bisset and Blakie}(2009)}]{Bisset2009b}
\bibinfo{author}{\bibfnamefont{R.~N.} \bibnamefont{Bisset}} \bibnamefont{and}
  \bibinfo{author}{\bibfnamefont{P.~B.} \bibnamefont{Blakie}},
  \bibinfo{journal}{Phys. Rev. A} \textbf{\bibinfo{volume}{80}},
  \bibinfo{pages}{035602} (\bibinfo{year}{2009}).

\bibitem[{\citenamefont{{Yang} and {Yang}}(1969)}]{Yang1969}
\bibinfo{author}{\bibfnamefont{C.~N.} \bibnamefont{{Yang}}} \bibnamefont{and}
  \bibinfo{author}{\bibfnamefont{C.~P.} \bibnamefont{{Yang}}},
  \bibinfo{journal}{Journal of Mathematical Physics}
  \textbf{\bibinfo{volume}{10}}, \bibinfo{pages}{1115} (\bibinfo{year}{1969}).

\bibitem[{\citenamefont{{Kheruntsyan} et~al.}(2010)\citenamefont{{Kheruntsyan},
  {Davis}, {Blakie}, {van Amerongen}, and {van Druten}}}]{Kheruntsyan2010}
\bibinfo{author}{\bibfnamefont{K.}~\bibnamefont{{Kheruntsyan}}},
  \bibinfo{author}{\bibfnamefont{M.~J.} \bibnamefont{{Davis}}},
  \bibinfo{author}{\bibfnamefont{P.~B.} \bibnamefont{{Blakie}}},
  \bibinfo{author}{\bibfnamefont{A.~H.} \bibnamefont{{van Amerongen}}},
  \bibnamefont{and} \bibinfo{author}{\bibfnamefont{N.~J.} \bibnamefont{{van
  Druten}}}, \bibinfo{journal}{Poster at ICAP}  (\bibinfo{year}{2010}).

\bibitem[{\citenamefont{Blakie et~al.}(2008)\citenamefont{Blakie, Bradley,
  Davis, Ballagh, and Gardiner}}]{Blakie2008}
\bibinfo{author}{\bibfnamefont{P.~B.} \bibnamefont{Blakie}},
  \bibinfo{author}{\bibfnamefont{A.~S.} \bibnamefont{Bradley}},
  \bibinfo{author}{\bibfnamefont{M.~J.} \bibnamefont{Davis}},
  \bibinfo{author}{\bibfnamefont{R.~J.} \bibnamefont{Ballagh}},
  \bibnamefont{and} \bibinfo{author}{\bibfnamefont{C.~W.}
  \bibnamefont{Gardiner}}, \bibinfo{journal}{Adv. Phys.}
  \textbf{\bibinfo{volume}{57}}, \bibinfo{pages}{363} (\bibinfo{year}{2008}).

\bibitem[{\citenamefont{Shvarchuck et~al.}(2002)\citenamefont{Shvarchuck,
  Buggle, Petrov, Dieckmann, Zielonkowski, Kemmann, Tiecke, von Klitzing,
  Shlyapnikov, and Walraven}}]{Shvarchuck2002}
\bibinfo{author}{\bibfnamefont{I.}~\bibnamefont{Shvarchuck}},
  \bibinfo{author}{\bibfnamefont{C.}~\bibnamefont{Buggle}},
  \bibinfo{author}{\bibfnamefont{D.~S.} \bibnamefont{Petrov}},
  \bibinfo{author}{\bibfnamefont{K.}~\bibnamefont{Dieckmann}},
  \bibinfo{author}{\bibfnamefont{M.}~\bibnamefont{Zielonkowski}},
  \bibinfo{author}{\bibfnamefont{M.}~\bibnamefont{Kemmann}},
  \bibinfo{author}{\bibfnamefont{T.~G.} \bibnamefont{Tiecke}},
  \bibinfo{author}{\bibfnamefont{W.}~\bibnamefont{von Klitzing}},
  \bibinfo{author}{\bibfnamefont{G.~V.} \bibnamefont{Shlyapnikov}},
  \bibnamefont{and} \bibinfo{author}{\bibfnamefont{J.~T.~M.}
  \bibnamefont{Walraven}}, \bibinfo{journal}{Phys. Rev. Lett.}
  \textbf{\bibinfo{volume}{89}}, \bibinfo{pages}{270404}
  (\bibinfo{year}{2002}).

\bibitem[{\citenamefont{Landau and Lifshitz}(1980)}]{LandauLifshitz_statmech1}
\bibinfo{author}{\bibfnamefont{L.~D.} \bibnamefont{Landau}} \bibnamefont{and}
  \bibinfo{author}{\bibfnamefont{E.~M.} \bibnamefont{Lifshitz}},
  \emph{\bibinfo{title}{Statistical Physics, Part 1.}}, vol.
  \bibinfo{volume}{Vol. 5 (3rd ed.)}
  (\bibinfo{publisher}{Butterworth-Heinemann}, \bibinfo{year}{1980}).

\bibitem[{\citenamefont{Cockburn et~al.}(2010)\citenamefont{Cockburn,
  Nistazakis, Horikis, Kevrekidis, Proukakis, and
  Frantzeskakis}}]{Cockburn2010}
\bibinfo{author}{\bibfnamefont{S.~P.} \bibnamefont{Cockburn}},
  \bibinfo{author}{\bibfnamefont{H.~E.} \bibnamefont{Nistazakis}},
  \bibinfo{author}{\bibfnamefont{T.~P.} \bibnamefont{Horikis}},
  \bibinfo{author}{\bibfnamefont{P.~G.} \bibnamefont{Kevrekidis}},
  \bibinfo{author}{\bibfnamefont{N.~P.} \bibnamefont{Proukakis}},
  \bibnamefont{and} \bibinfo{author}{\bibfnamefont{D.~J.}
  \bibnamefont{Frantzeskakis}}, \bibinfo{journal}{Phys. Rev. Lett.}
  \textbf{\bibinfo{volume}{104}}, \bibinfo{pages}{174101}
  (\bibinfo{year}{2010}).

\bibitem[{\citenamefont{{Weiler} et~al.}(2008)\citenamefont{{Weiler}, {Neely},
  {Scherer}, {Bradley}, {Davis}, and {Anderson}}}]{Weiler2008}
\bibinfo{author}{\bibfnamefont{C.~N.} \bibnamefont{{Weiler}}},
  \bibinfo{author}{\bibfnamefont{T.~W.} \bibnamefont{{Neely}}},
  \bibinfo{author}{\bibfnamefont{D.~R.} \bibnamefont{{Scherer}}},
  \bibinfo{author}{\bibfnamefont{A.~S.} \bibnamefont{{Bradley}}},
  \bibinfo{author}{\bibfnamefont{M.~J.} \bibnamefont{{Davis}}},
  \bibnamefont{and} \bibinfo{author}{\bibfnamefont{B.~P.}
  \bibnamefont{{Anderson}}}, \bibinfo{journal}{Nature}
  \textbf{\bibinfo{volume}{455}}, \bibinfo{pages}{948} (\bibinfo{year}{2008}),
  \eprint{0807.3323}.

\bibitem[{\citenamefont{Damski and Zurek}(2010)}]{Damski2010}
\bibinfo{author}{\bibfnamefont{B.}~\bibnamefont{Damski}} \bibnamefont{and}
  \bibinfo{author}{\bibfnamefont{W.~H.} \bibnamefont{Zurek}},
  \bibinfo{journal}{Phys. Rev. Lett.} \textbf{\bibinfo{volume}{104}},
  \bibinfo{pages}{160404} (\bibinfo{year}{2010}).

\bibitem[{\citenamefont{{Das} et~al.}(2011)\citenamefont{{Das}, {Sabbatini},
  and {Zurek}}}]{Das2011}
\bibinfo{author}{\bibfnamefont{A.}~\bibnamefont{{Das}}},
  \bibinfo{author}{\bibfnamefont{J.}~\bibnamefont{{Sabbatini}}},
  \bibnamefont{and} \bibinfo{author}{\bibfnamefont{W.~H.}
  \bibnamefont{{Zurek}}} (\bibinfo{year}{2011}), \eprint{arXiv:1102.5474}.

\bibitem[{\citenamefont{Cockburn and Proukakis}(2009)}]{Cockburn2009}
\bibinfo{author}{\bibfnamefont{S.~P.} \bibnamefont{Cockburn}} \bibnamefont{and}
  \bibinfo{author}{\bibfnamefont{N.~P.} \bibnamefont{Proukakis}},
  \bibinfo{journal}{Las. Phys.} \textbf{\bibinfo{volume}{19}},
  \bibinfo{pages}{558} (\bibinfo{year}{2009}).

\bibitem[{\citenamefont{{Zaremba} et~al.}(1999)\citenamefont{{Zaremba},
  {Nikuni}, and {Griffin}}}]{Zaremba1999}
\bibinfo{author}{\bibfnamefont{E.}~\bibnamefont{{Zaremba}}},
  \bibinfo{author}{\bibfnamefont{T.}~\bibnamefont{{Nikuni}}}, \bibnamefont{and}
  \bibinfo{author}{\bibfnamefont{A.}~\bibnamefont{{Griffin}}},
  \bibinfo{journal}{J. Low Temp. Phys.} \textbf{\bibinfo{volume}{116}},
  \bibinfo{pages}{277} (\bibinfo{year}{1999}).

\bibitem[{\citenamefont{Andersen et~al.}(2002)\citenamefont{Andersen,
  Al~Khawaja, and Stoof}}]{Andersen2002}
\bibinfo{author}{\bibfnamefont{J.}~\bibnamefont{Andersen}},
  \bibinfo{author}{\bibfnamefont{U.}~\bibnamefont{Al~Khawaja}},
  \bibnamefont{and} \bibinfo{author}{\bibfnamefont{H.}~\bibnamefont{Stoof}},
  \bibinfo{journal}{Phys. Rev. Lett.} \textbf{\bibinfo{volume}{88}},
  \bibinfo{pages}{070407} (\bibinfo{year}{2002}).

\bibitem[{\citenamefont{Al~Khawaja et~al.}(2002)\citenamefont{Al~Khawaja,
  Andersen, Proukakis, and Stoof}}]{AlKhawaja2002}
\bibinfo{author}{\bibfnamefont{U.}~\bibnamefont{Al~Khawaja}},
  \bibinfo{author}{\bibfnamefont{J.~O.} \bibnamefont{Andersen}},
  \bibinfo{author}{\bibfnamefont{N.~P.} \bibnamefont{Proukakis}},
  \bibnamefont{and} \bibinfo{author}{\bibfnamefont{H.~T.~C.}
  \bibnamefont{Stoof}}, \bibinfo{journal}{Phys. Rev. A}
  \textbf{\bibinfo{volume}{66}}, \bibinfo{pages}{013615}
  (\bibinfo{year}{2002}), \bibinfo{note}{erratum: Phys. Rev. A {\bf 66},
  059902(E) (2002).}

\end{thebibliography}

\end{document}